\documentclass[12pt,a4paper]{article}

\usepackage{mathptmx}          % Times text + Times math, matches the figures

\usepackage{amsmath,amssymb,amsfonts}
\usepackage{graphicx}
\usepackage{bm}
\usepackage{hyperref}
\usepackage{booktabs}
\usepackage{xcolor}
\usepackage{tikz}
\usepackage{quantikz}
\usetikzlibrary{patterns,positioning,fit,backgrounds}
\usepackage[T1]{fontenc}
\usepackage{lmodern}
\usepackage{amsmath,amssymb,amsthm,mathtools,bm}
\usepackage{microtype}
\usepackage{geometry}
\usepackage{hyperref}
\usepackage{cite}
\usepackage{booktabs}
\usepackage{xcolor}
\usepackage{graphicx}
\usepackage{float}
\hypersetup{colorlinks=true,linkcolor=blue!55!black,citecolor=blue!55!black,urlcolor=blue!55!black}

\newcommand{\acknowledgments}{\section*{Acknowledgments}}

\hypersetup{colorlinks=true,linkcolor=blue,citecolor=blue,urlcolor=blue}

\newcommand{\Z}{\mathbb{Z}}

\providecommand{\ket}[1]{|#1\rangle}
\newcommand{\ev}[1]{\langle#1\rangle}
\usepackage{titling}
\pretitle{\begin{center}\fontsize{16}{18}\selectfont\bfseries}
\posttitle{\par\end{center}}
\title{$p$-Form Gauge Dynamics and Digital Quantum Simulation\\[0.2cm]
-- Flux and Cosmological Constant Neutralization --}
\author{\sc Soo-Jong Rey\\[0.2cm]
{\sl Kwangwoon University}\\
{\sl Seoul, Korea}}
\date{\tt sjrey@kw.ac.kr}

\begin{document}

\maketitle

\begin{abstract}
I develop a Hamiltonian framework for $\mathbb Z_k$ $p$-form gauge
fields on arbitrary oriented cell complexes in arbitrary dimensions.
Gauge qudits are defined by $p$-cells, charged boundary qudits by
$(p-1)$-cells, Gauss-law generators by boundary map $\partial_p$,
and magnetic checks by $\partial_{p+1}$.  The same cellular structure produces local dressed Wilson operators, and at $k=2$ a Calderbank--Shor--Steane check complex relevant to quantum error correction.
I then specialize to $p=2, k=2$, where the magnetic 3-cell term is absent and the one-form Gauss-law can be solved exactly. The physical Hilbert space is parameterized by plaquette electric-flux variables, while the link configuration is reconstructed as the dynamical boundary of the evolving flux domains.  The reduced Hamiltonian is an Ising-type plaquette model, where its local transverse-field term is the physical
image of the boundary-dressed Wilson operator
$\sigma_p^z\prod_{\ell\in\partial p}\tau_\ell^z$.
A tube--cap quench compares two initial flux fillings with the same initial
boundary loops.  Exact diagonalization on $4\times4$, $6\times4$, and
$5\times5$ tori finds that the cap loses $20$--$37\%$ of its occupied-flux area, while the tube remains nearly pinned.  
A finite-size scaling locates a dynamical crossover of tension-to-density ratio near $(m/\varepsilon_E)_c\simeq1.89$.  
The unreduced plaquette-plus-link encoding provides local Gauss-law checks and a direct digital implementation, while 
the reduced plaquette-only Hamiltonian supplies the exact benchmark.  The result places the specific top-form discharge and the cosmological constant neutralization calculation inside a general higher-form
Hamiltonian and coding framework.
\end{abstract}

\section{Introduction}

\subsection{Exact dynamics of Gauss-law reduced two-form lattice model}
\label{sec:intro_neutralization}

In this paper, built upon my previous series of works on Kalb-Ramond gauge theory
\cite{KalbRamond,ReyHiggs1989,ReyConfining1991,ReySugino2010}, I develop two levels.  Section~\ref{sec:general-higher-form} first formulates a compact $p$-form gauge Hamiltonian of gauge group $G = \mathbb Z_k$ on an arbitrary oriented cell complex in arbitrary dimensions.  The boundary maps define the Gauss-law generators, magnetic
operators, and dressed Wilson operators. This formulation separates the general higher-form operator algebra from the particular top-form model (used below in the numerical calculation for flux discharge and cosmological constant neutralization) and exposes the sparse Gauss-law-check structure used in gauge-symmetry-adapted quantum error detection (QEC) and gauge-covariant simulation~\cite{Wiebe2023,Wiebe2026}. 

The backbone of the present work remains the exact real-time evolution of the specialized Gauss-law-projected, $G = \Z_2$ Hamiltonian. Solving the one-form Gauss-law
eliminates the link variables from the independent Hilbert space and leaves one
qubit per plaquette. The eliminated links remain derived
observables,
\[
  \tau^x_\ell=\sigma^x_{p_L}\sigma^x_{p_R}.
\]
Consequently, the closed link loops are the time-dependent domain-wall
boundaries of the evolving plaquette-flux configuration. They are dynamical in
the same precise sense that a Gauss-law-reconstructed electric field is dynamical
in reduced Schwinger-model Hamiltonian, although they are not independent
matter coordinates.

The quench begins from two distinct plaquette-flux product states having the same initial
boundary loops and different fillings of the torus. The tube state occupies the
strip between the loops; the cap state occupies the complementary strip. The
reduced Hamiltonian then evolves the complete physical Hilbert space. No
boundary geometry is held fixed after preparation. Local plaquette flips alter
the adjacent derived boundary links automatically, so boundary loops appear,
deform, merge, and disappear as consequences of the flux dynamics.

Exact diagonalization on $4\times4$, $6\times4$, and $5\times5$ tori provides
the principal results. The continuum and unreduced lattice formulation explain
the origin of the reduced Hamiltonian, the gauge-invariant operator algebra,
and the observables reconstructed from each exact diagonalization trajectory. The digital-circuit
construction implements the same dynamics in a redundant plaquette-plus-link
encoding and permits Gauss-law-sector verification at readout.

The measured quantities are the occupied-flux area
$M(t)=\sum_p(1-\sigma_p^x)/2$ and the derived boundary length
$L(t)=\sum_\ell(1-\tau_\ell^x)/2$. The cap--tube contrast isolates the
relaxation of two inequivalent initial flux fillings. The connection to Brown--Teitelboim cosmological constant neutralization~\cite{BrownTeitelboim1987,BrownTeitelboim1988}  is used at the level of top-form flux discharge; the present
model contains neither gravity nor an independent continuum string field.

\begin{figure}[!htbp]
\centering
\includegraphics[width=0.35\linewidth]{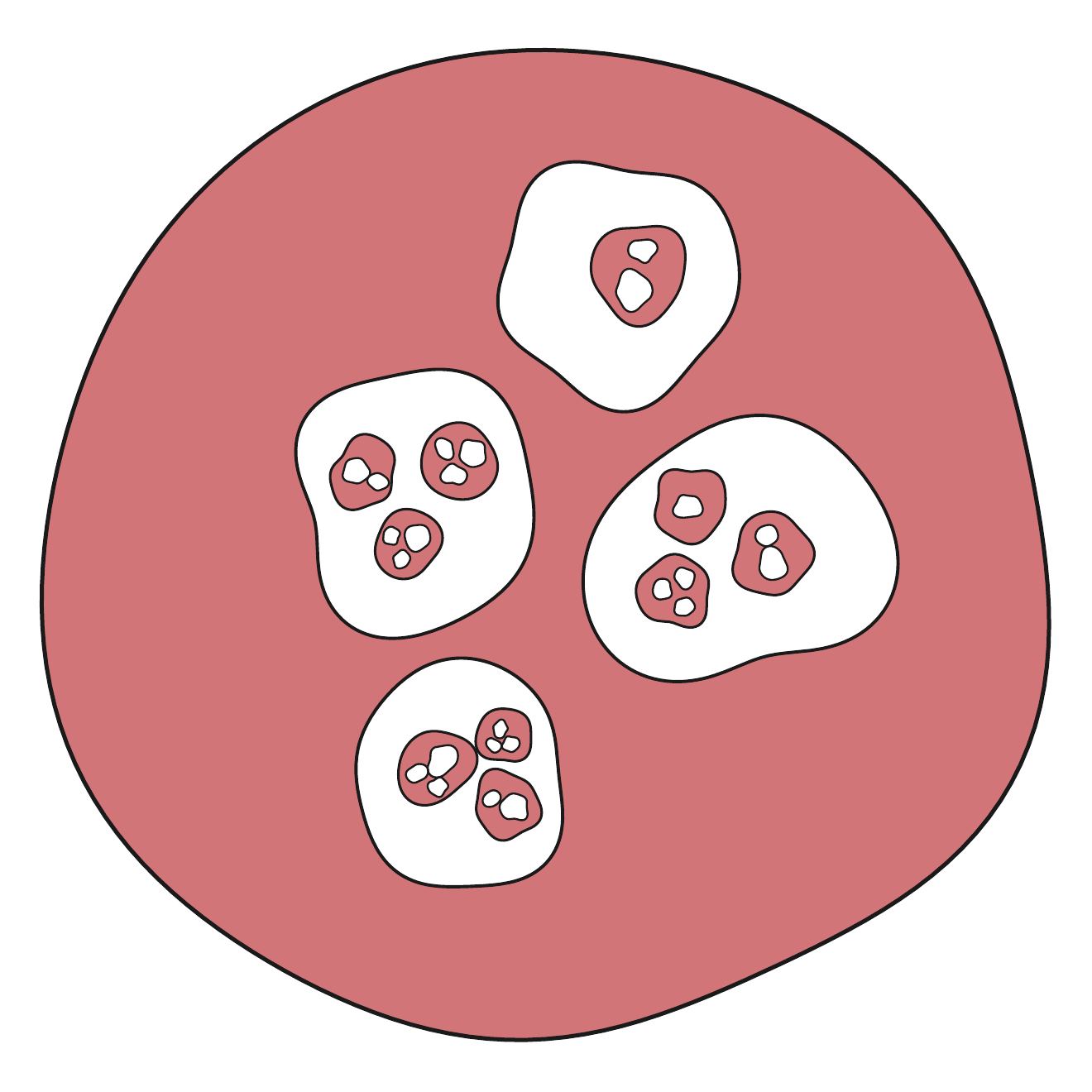}
\caption{The late-time configuration anticipated for the
post-discharge string-Higgs phase. The outer red region carries
top-form electric flux; white islands are the dual vacuum. Each
filled--empty interface is a closed matter string ($\tau^x_\ell = -1$
along the curve). Repeated nucleation events at successively smaller
scales results from non-equilibrium relaxation dynamics and generate a recursive nested-island structure---filled regions
inside empty regions inside filled regions. The protocol of this paper resolves the dynamics by which such
configurations are produced from a uniform high-flux initial state
(Sec.~\ref{sec:ED}); the equilibration sector accessible at late times
is set by the tension-to-energy-density ratio $m/\varepsilon_E$, with a sharp crossover at
$(m/\varepsilon_E)_c \approx 1.89$ (Sec.~\ref{sec:m-eps-sweep}).}
\label{fig:percolation_nested}
\end{figure}

\subsection{Quantum simulation of lattice gauge theories: state of the art}
\label{sec:intro_lgt}

Real-time quantum simulation of lattice gauge theories has progressed rapidly. On the theory side, analog and digital protocols have been worked out for
$U(1)$, $\Z_N$, and non-abelian models with fermionic or scalar matter
\cite{BanulsEPJD2020,DiMeglio2024}. On the hardware side, (1+1)-dimensional string formation and breaking---the Schwinger mechanism of $q\bar q$ pair
production inside an electric flux tube---has been the proving ground.
Hebenstreit, Berges, and Gelfand analyzed real-time string-breaking
dynamics in 1+1D QED, but  classically~\cite{HebenstreitBergesGelfand2013}. They
established the two-stage process of pair creation and charge separation
and multiple string breaking when the static charges
are replaced by dynamical charges flying apart. Tensor-network studies by
Pichler, Dalmonte, Rico, Zoller, and Montangero~\cite{Pichler2016} mapped
out the real-time U(1) string-breaking dynamical state diagram with
dynamical matter. Subsequent tensor-network work extended this to
\emph{multi-particle} processes: Rigobello \emph{et al.}\
\cite{Rigobello2021} studied entanglement generation in $(1+1)$-dimensional QED scattering
processes; Papaefstathiou, Knolle, and Ba\~nuls \cite{PapaefstathiouBanuls2025}
simulated meson--meson collisions in the Schwinger model and used the
two-point \emph{electric flux correlator} $C_\ell^{mn}(t) =
\langle E_m(t) E_n(t)\rangle - \langle E_m(t)\rangle\langle E_n(t)\rangle$
as the diagnostic distinguishing elastic from inelastic channels;
Belyansky \emph{et al.}\ \cite{BelyanskyDavoudiGorshkov2024} constructed
multi-particle wave packets via uniform matrix product states in the
bosonized Schwinger model and identified inelastic quark and meson
production, meson disintegration, and dynamical string formation/breaking. Lin, Luo, Yao and Shanahan solved the Lindblad master equation of the Schwinger model and studied open quantum system effects to string breaking and recombination~\cite{Lin2024}. Going beyond qubits, Ale \emph{et al.}\ extended the digital simulation of $(2+1)$-dimensional QED to qumons~\cite{Ale2026}. 

Trapped-ion hardware demonstrations of the same circle of physics include
the Schwinger-model pair creation experiment of Nguyen \emph{et al.}\
\cite{Nguyen2022Schwinger}, the meson wave-packet preparation and collision
of Davoudi, Hsieh, and Kadam on the Quantinuum H1-1 and IonQ Forte platforms
\cite{DavoudiHsiehKadam2024, DavoudiHsiehKadam2025}, the meson-scattering
spin-simulator protocol of Bennewitz \emph{et al.}\ \cite{Bennewitz2025}, and
the IBM hadron-scattering experiment of Schuhmacher \emph{et al.}\
\cite{Schuhmacher2025}. In $(2+1)$ dimensions, the analogous one-form program has just
produced three recent hardware demonstrations: $\Z_2$ Higgs string
dynamics on IBM heavy-hex \cite{CobosZ2Higgs}, quench-induced string
formation and breaking on Google Sycamore \cite{Cochran2025Nature}, and
$U(1)$ string breaking on Rydberg atom arrays in the Kagome geometry
\cite{GonzalezCuadra2025Nature}. All of these are one-form gauge theories:
electric charges are point particles, flux is supported on hadronic
strings, and the flux tube between a $q\bar q$ pair has a unique
topological type.

\subsection{The Kalb--Ramond two-form setting}
\label{sec:intro_kr}

The general construction is specialized to $p=2$, $k=2$ and two spatial dimensions.  Its
continuum origin is the compact $(2+1)$-dimensional version of Kalb--Ramond gauge dynamics~\cite{KalbRamond}, firsts studied in the context of string theory where the pull-back of the two-form potential is coupled to the worldsheet of charged string matter. The canonical
electric field $E^{ij}=H^{0ij}$ is a spatial two-form and therefore lives on
plaquettes of the spatial lattice. A one-form Gauss-law is attached to links.
Because $p=d$, no spatial $(p+1)$-cells exist and the magnetic term of the
general Hamiltonian vanishes.  In the hard-core $\Z_2$ truncation used here,
the Gauss-law can be solved exactly, leaving the plaquette electric variables
as a complete set of physical coordinates.

The unreduced encoding realizes the one-form Gauss-law constraints  as sparse local checks. the present protocol uses these checks for gauge-sector verification and post-selection, providing the detection layer of gauge-symmetry-adapted QEC rather than a complete active recovery scheme. 

The unreduced representation retains one qubit on each plaquette and one on
each link. Its elementary gauge-invariant off-diagonal operator is
\[
  W_p=\sigma_p^z\prod_{\ell\in\partial p}\tau_\ell^z .
\]
After projection, $W_p$ acts simply as $\sigma_p^z$ on the physical plaquette basis. 
This identity is the central bridge between the field-theory,
the local quantum circuit, and the exact diagonalization. 
The link occupation measured in the
unreduced encoding is not an additional independent matter variable; 
it is the
domain-wall observable determined by adjacent plaquette fluxes.

The continuum description of charged strings and Wilson surfaces identifies the
gauge symmetry and local invariant operator. 
The model I study hereafter is the
exact $\Z_2$-Gauss-law-reduced theory. This is not yet a full-fledged second-quantized string field theory -- it belongs to enlarged $G = \Z_k (k \ge 3)$, rotor or bosonic-link models.

\subsection{Structure of exact diagonalization and quantum circuit formulations}

Five features organize the calculation. 
\begin{enumerate}
\item[(i)] \emph{Exact Gauss-law reduction.} The link constraints
$\mathcal G_\ell=\tau_\ell^x\sigma_{p_L}^x\sigma_{p_R}^x=1$ determine every link
occupation from neighboring plaquette fluxes. 
The physical Hilbert space has
dimension $2^{N_p}$ and is represented exactly by plaquette qubits.

\item[(ii)] \emph{Derived closed boundaries.} Links with
$\tau_\ell^x=-1$ are domain walls of the plaquette-flux pattern. They form
closed loops and evolve whenever the plaquette configuration evolves. They are
dynamical observables, not fixed external sources and not independent matter
coordinates.

\item[(iii)] \emph{Local gauge-invariant transition.} In the redundant
encoding, $\mathcal W_p=\sigma_p^z\prod_{\ell\in\partial p}\tau_\ell^z$ preserves all
Gauss-law constraints. In the reduced Hilbert space it becomes the single-site
plaquette flip $\sigma_p^z$ used in ED.

\item[(iv)] \emph{Same-initial-boundary quench.} The tube and cap states share
the same initial domain-wall loops but occupy complementary plaquette regions.
The subsequent loops are not constrained to remain equal to the initial ones.

\item[(v)] \emph{Two equivalent implementations.} The exact diagonalization uses the reduced
plaquette-only Hamiltonian. The digital quantum circuit may use the redundant
plaquette-plus-link encoding, where Gauss-law checks diagnose leakage from the physical subspace.
\end{enumerate}

The headline observable is $M_{\rm cap}(t)-M_{\rm tube}(t)$ between two ED
trajectories initialized with different plaquette fillings. The boundary length
$L(t)$ is reconstructed from the same trajectories through Gauss-law. The
remainder of the paper derives the reduced model, specifies the circuit
representation, and reports the exact diagonalization results. 

\section{General higher-form Hamiltonian and the two-form specialization}
\label{sec:continuum}

\subsection{Compact $p$-form gauge theory on an oriented cell complex}
\label{sec:general-higher-form}

Let $D=(d+1)$ be the spacetime dimension and let $\mathcal C$ be a finite, oriented $d$-dimensional {\sl spatial} cell complex.  Denote the set of oriented $r$-cells by $\mathcal C_r$ and the cellular boundary maps by
\[
  C_{p+1}(\mathcal C;\mathbb Z_k)
  \xrightarrow{\,\partial_{p+1}\,}
  C_p(\mathcal C;\mathbb Z_k)
  \xrightarrow{\,\partial_p\,}
  C_{p-1}(\mathcal C;\mathbb Z_k),
  \qquad
  \partial_p\partial_{p+1}=0.
\]
For a $(p-1)$-cell $a$, a $p$-cell $f$, and a $(p+1)$-cell $c$, write
\[
  s_{af}=\langle a,\partial_p f\rangle,
  \qquad
  s_{fc}=\langle f,\partial_{p+1}c\rangle,
\]
with $s\in\{0,\pm1\}$ for an ordinary oriented cellulation.  The construction
below depends only on these sparse incidence matrices and therefore applies to regular lattices, simplicial complexes, hyperbolic cellulations, and general bounded-degree complexes.

Place one $\mathbb Z_k$ gauge qudit on every $p$-cell $f\in\mathcal C_p$ and,
for the minimal matter-coupled model, one charged boundary qudit on every
$(p-1)$-cell $a\in\mathcal C_{p-1}$.  Let $(Z_f,X_f)$ and $(z_a,x_a)$ denote clock and shift operators at each respective cells satisfying
\[
  Z_fX_f=\omega X_fZ_f,
  \qquad
  z_ax_a=\omega x_az_a,
  \qquad
  \omega=e^{2\pi i/N}.
\]
The local gauge generator attached to $a$ is
\begin{equation}
  \mathcal G_a=x_a\prod_{f\in\mathcal C_p}X_f^{s_{af}}
  \label{eq:general-gauss}
\end{equation}
Physical states satisfy $G_a|\psi\rangle=+|\psi\rangle$ for every
$a\in\mathcal C_{p-1}$.  
Removing the matter qudits gives the pure-gauge Gauss
operator
\[
  \mathcal A_a=\prod_fX_f^{s_{af}}.
\]

The elementary magnetic or curvature operator is attached to a
$(p+1)$-cell:
\begin{equation}
  \mathcal B_c=\prod_{f\in\mathcal C_p}Z_f^{s_{fc}}.
  \label{eq:general-magnetic}
\end{equation}
Gauge invariance follows directly from
$\partial_p\partial_{p+1}=0$, which gives
$[A_a,\mathcal B_c]=0$ for all $a,c$.  When $p=d$, no spatial
$(p+1)$-cells exist and the magnetic term is absent.  
This top-form limit is
precisely the situation I use later in the $(2+1)$-dimensional calculation.

A local open Wilson $p$-cell must be dressed by charged matter on its
boundary.  The oriented gauge-invariant operator is
\begin{equation}
  \mathcal W_f
  =Z_f\prod_{a\in\mathcal C_{p-1}}z_a^{-s_{af}},
  \qquad
  [\mathcal W_f,G_a]=0.
  \label{eq:general-wilson}
\end{equation}
It changes the gauge flux on $f$ together with the charged
$(p-1)$-dimensional boundary carried by $\partial f$.  A second local invariant,
used in the present truncation, is the diagonal boundary-charge interaction
\begin{equation}
  \mathcal K_f=\prod_{a\in\mathcal C_{p-1}}x_a^{s_{af}}.
  \label{eq:general-kappa}
\end{equation}

A compact Hamiltonian containing the electric, magnetic, 
boundary-energy, Wilson, and diagonal boundary-interaction terms is
\begin{equation}
{
\begin{aligned}
 H_{p,d}^{(k)}={}&
 \frac{\varepsilon_E}{4}\sum_{f\in\mathcal C_p}
     \left(2-X_f-X_f^\dagger\right)
 -\frac{J_B}{2}\sum_{c\in\mathcal C_{p+1}}
     \left(\mathcal B_c+\mathcal B_c^\dagger\right) \\
 &+\frac{m}{4}\sum_{a\in\mathcal C_{p-1}}
     \left(2-x_a-x_a^\dagger\right)
 -\frac{t}{2}\sum_{f\in\mathcal C_p}
     \left(\mathcal W_f+\mathcal W_f^\dagger\right) \\
 &-\frac{\kappa}{2}\sum_{f\in\mathcal C_p}
     \left(\mathcal K_f+\mathcal K_f^\dagger\right).
\end{aligned}}
\label{eq:general-H}
\end{equation}
Every term commutes with every $\mathcal G_a$.  
I stress that Eq.~\eqref{eq:general-H} is the 
minimal family rather than a unique universal Hamiltonian: additional local
class functions of the electric variables, multi-cell interactions, or
independent matter dynamics may be added without changing the constraint
algebra.  
Its purpose here is to expose the dimension-independent structure
from which the exact diagonalization model follows by a direct specialization.

\subsection{Gauss-law constraints as a higher-form check complex}
\label{sec:qec-structure}

For $k=2$ and in the pure-gauge sector, the operators
\[
  S_a^X=\mathcal A_a=\prod_fX_f^{s_{af}},
  \qquad
  S_c^Z=\mathcal B_c=\prod_fZ_f^{s_{fc}}
\]
form a Calderbank--Shor--Steane(CSS) stabilizer algebra.  Their commutation is the cellular identity
$\partial_p\partial_{p+1}=0$.  On a bounded-degree complex, every check has
bounded weight and every qubit participates in a bounded number of checks, so
this is a quantum low-density parity-check (qLDPC) code structure.  The encoded homological sectors are governed
by
\[
  H_p(\mathcal C;\mathbb Z_2)
  =\ker\partial_p/\operatorname{im}\partial_{p+1},
\]
while the dual logical operators are represented by $p$-cocycles.  Code rate
and distance are therefore properties of the chosen complex, not of the form
degree alone.

Three coding problems must be separated. First, in gauge-covariant Hamiltonian simulation only the Gauss operators ${\cal A}_a$ or ${\cal G}_a$ are enforced, while the magnetic operators remain dynamical. Second, treating both ${\cal A}_a$ and ${\cal B}_c$ as stabilizers defines a static homological CSS memory and freezes the corresponding local gauge dynamics. Third, a subsystem construction may retain local dynamics on gauge degrees of freedom while protecting selected homological sectors. The compatibility of sparse checks, growing dressed distance, and low-weight Wilson dynamics is the central open problem.

The matter-coupled generators $\mathcal G_a=x_a\mathcal A_a$ retain sparse local syndrome checks. Interpreting the gauge-invariant sector as a code space follows the Gauss-law error-correction framework of Refs.~\cite{Wiebe2023,Wiebe2026}. In this work, these checks are used for leakage detection and final-time post-selection, rather than for active fault-tolerant recovery. 
They do not by themselves guarantee a growing-distance quantum
code: the added matter qudits can support low-weight gauge-preserving errors.
Thus two distinct QEC usages arise from the same formalism.  The pure-gauge three-term complex supplies a homological CSS/qLDPC code candidate, whereas the matter-coupled model supplies a redundant gauge encoding with local Gauss
syndromes. In the present paper I use the latter for circuit verification and the
former only to display the general extension.  Constructing finite-rate,
growing-distance families requires a separate choice of cell complex and lies beyond the exact diagonalization calculation.

The dimensional specialization matters.  For a $p$-form in $d=p$ spatial
dimensions, $\mathcal C_{p+1}$ is empty, so the local $Z$-check family
$\mathcal B_c$ is absent.  The resulting top-form Hamiltonian still has local Gauss-law checks, but it is not by itself a complete homological qLDPC code.  For
$d\ge p+1$, both check families are present and the full chain complex becomes available.

So, in the pure $p=2,d=3$ cubic model, the Gauss-law checks obey the local relations $\prod_{e\ni v}{\cal A}_e=1$. Consequently, face-supported Z faults generate closed loop syndromes, and the vertex relations provide meta-checks for noisy syndrome extraction. The matter-coupled top-form model calculated in the rest of the present paper does not possess this same intrinsic redundancy: its ${\cal G}_\ell$ generators are independent because each contains a unique link operator. The current post-selection protocol therefore performs gauge-sector verification rather than single-shot error correction.

\subsection{Continuum two-form gauge theory in $2+1$ dimensions}

The numerical model is the specialization
\[
  p=2,
  \qquad d=2,
  \qquad D=3,
  \qquad k=2.
\]
Its continuum origin is the $U(1)$ Kalb--Ramond two-form gauge field
$B_{\mu\nu}(x)=-B_{\nu\mu}(x)$ with field strength
\begin{equation}
  H_{\mu\nu\rho}
  =\partial_\mu B_{\nu\rho}
  +\partial_\nu B_{\rho\mu}
  +\partial_\rho B_{\mu\nu}.
\end{equation}
The pure gauge dynamics action is
\begin{equation}
  S_\mathrm{KR}
  =-\frac{1}{12}\int d^D x\,H_{\mu\nu\rho}H^{\mu\nu\rho}.
\end{equation}
The gauge redundancy is
\begin{equation}
  B_{\mu\nu}\to B_{\mu\nu}
  +\partial_\mu\lambda_\nu-\partial_\nu\lambda_\mu,
\end{equation}
with the gauge-for-gauge redundancy
$\lambda_\mu\to\lambda_\mu+\partial_\mu\alpha$.  A massless $p$-form in $D$
spacetime dimensions has $\binom{D-2}{p}$ local polarizations.  For $p=2$ and
$D=3$ this vanishes.  The Hodge dual $\star H$ is a scalar field
strength, and the source-free equation of motion fixes it to a spacetime
constant, the source of the cosmological constant.

In the continuum parent theory, the electrically charged object is a string.
A string worldsheet $\Sigma$ couples through
$\exp(iq\int_\Sigma B)$.  On a fixed Hamiltonian slice, an open Wilson surface
supported on a spatial region $R$ transforms by a phase on its boundary
$C=\partial R$ and therefore requires charged boundary dressing.  The
spacelike support $R_t$ of this equal-time operator is not part of the timelike string
worldsheet.  A time-dependent boundary $C_t$ sweeps out
$\Sigma=\bigcup_tC_t$, while the family of electric-flux regions $R_t$ fills a spacetime three-chain.

Temporal gauge $B_{0i}=0$ leaves the spatial component $B_{ij}$.  In two
spatial dimensions, $B_{12}$ is the only independent component.  Its canonical
electric field is the spatial two-form $E^{ij}=H^{0ij}=\epsilon^{ij}E$.  The
Gauss-law is
\begin{equation}
  \partial_iE^{ij}=\rho^j_\mathrm{string},
  \qquad\text{or}\qquad
  \epsilon^{ij}\partial_iE=\rho^j_\mathrm{string}.
\end{equation}
The purely spatial curvature $H_{ijk}$ requires three spatial indices and is
absent.  This is the continuum counterpart of the missing magnetic term in
Eq.~\eqref{eq:general-H} when $p=d=2$.

\subsection{Square-lattice realization and $\mathbb Z_2$ truncation}

On the square lattice, the general cells become
\[
  f\in\mathcal C_2\longleftrightarrow p\ \text{(plaquette)},
  \qquad
  a\in\mathcal C_1\longleftrightarrow \ell\ \text{(link)}.
\]
The compact gauge variable $B_p\in U(1)$ and its conjugate electric field $E_p$
live on plaquettes, equivalently on dual-lattice sites.  The one-form gauge
parameter $\eta_\ell$ lives on links and acts as
\begin{equation}
  B_p\to B_p+\sum_{\ell\in\partial p}s_{\ell p}\eta_\ell.
\end{equation}
A charged link phase $\phi_\ell$ transforms as
$\phi_\ell\to\phi_\ell+\eta_\ell$.  With
$U_p=e^{iB_p}$ and $V_\ell=e^{i\phi_\ell}$, the elementary oriented Wilson
operator is
\[
  U_p\prod_{\ell\in\partial p}V_\ell^{-s_{\ell p}}.
\]

For $N=2$, place Pauli operators $\sigma_p^{x,y,z}$ on plaquettes and
$\tau_\ell^{x,y,z}$ on links.  The convention is
\[
  \sigma_p^z:\ \text{two-form potential},
  \qquad
  \sigma_p^x:\ \text{plaquette electric field},
\]
\[
  \tau_\ell^z:\ \text{charged boundary phase},
  \qquad
  \tau_\ell^x:\ \text{boundary-charge variable}.
\]
The eigenvalue $\sigma_p^x=-1$ marks an occupied electric-flux plaquette, and
$\tau_\ell^x=-1$ marks an activated boundary link.  Orientation signs disappear
because inverse and direct $\mathbb Z_2$ operators coincide.

\subsection{One-form Gauss-law and exact reduction}
\label{sec:gauss}

Equation~\eqref{eq:general-gauss} becomes
\begin{equation}
\label{eq:Gauss}
  \mathcal G_\ell=\tau_\ell^x\prod_{p\ni\ell}\sigma_p^x.
\end{equation}
Physical states satisfy
\[
  \mathcal G_\ell|\psi_\mathrm{phys}\rangle=|\psi_\mathrm{phys}\rangle
  \qquad\forall\ell.
\]
Every $\mathcal G_\ell$ contains a unique link operator $\tau_\ell^x$, so the
constraints are independent and
\[
  \dim\mathcal H_\mathrm{phys}=2^{N_p}.
\]
For an interior link between plaquettes $p_L$ and $p_R$,
\begin{equation}
  \tau_\ell^x=\sigma_{p_L}^x\sigma_{p_R}^x.
  \label{eq:gauss-reduction}
\end{equation}
Thus the plaquette electric fields form a complete set of physical coordinates.
The link pattern is reconstructed from them at every time.  Links with
$\tau_\ell^x=-1$ are domain walls of the instantaneous plaquette-flux
configuration; they are dynamical derived observables, not fixed external
sources and not independent matter variables.

The same statement on the dual lattice reads
\[
 { \mathcal G}_{\ell^\ast}
  =\tau_{\ell^\ast}^x \ \sigma_{p_L^\ast}^x \ \sigma_{p_R^\ast}^x=1.
\]
In the $U(1)$ theory it is the oriented difference equation
\[
  \sum_{p\supset\ell}s_{\ell p}E_p=\rho_\ell^\mathrm{string}.
\]
At every interior vertex,
$\prod_{\ell\supset v}\tau_\ell^x=1$, so the activated boundary links form
closed one-cycles.  This exact Gauss-law reduction is the representation used in the exact diagonalization calculation below.

\subsection{Gauge-invariant operators and the specialized Hamiltonian}

The measured electric-flux area and derived boundary length are
\[
  M=\sum_p\frac{1-\sigma_p^x}{2},
  \qquad
  L=\sum_\ell\frac{1-\tau_\ell^x}{2}.
\]
The specialized diagonal boundary interaction is
\[
  \mathcal K_p=\prod_{\ell\in\partial p}\tau_\ell^x,
\]
which becomes
${\mathcal K}_p=\prod_{p'\sim p}\sigma_{p'}^x$ after Gauss-law reduction.

For a connected plaquette region $R$ with boundary $\gamma=\partial R$, the
$\mathbb Z_2$ Wilson operator is
\begin{equation}
\label{eq:Wilson}
  \mathcal W[R,\gamma]
  =\left(\prod_{p\in R}\sigma_p^z\right)
   \left(\prod_{\ell\in\gamma}\tau_\ell^z\right).
\end{equation}
On a boundary link, the $\tau_\ell^z$ and adjacent $\sigma_p^z$ factors each
anticommute once with $\mathcal G_\ell$, so the two signs cancel.  Interior links meet
two plaquettes in $R$ and are likewise invariant.  The elementary operator is
\[
  \mathcal W_p=\sigma_p^z\prod_{\ell\in\partial p}\tau_\ell^z.
\]
It flips one plaquette electric field and the four adjacent derived boundary
links while preserving every Gauss-law constraint.  In the reduced physical basis,
$\mathcal W_p$ acts as a single plaquette flip.

The Hamiltonian used throughout the remainder of the paper is the
$p=d=k=2$ specialization of Eq.~\eqref{eq:general-H}, $H^{(2)}_{2,2} := H$:
\begin{equation}
\label{eq:H}
{
\begin{aligned}
  H={}&\frac{\varepsilon_E}{2}\sum_p(1-\sigma_p^x)
      -t\sum_p\sigma_p^z\prod_{\ell\in\partial p}\tau_\ell^z \\
     &-\kappa\sum_p\prod_{\ell\in\partial p}\tau_\ell^x
      +\frac{m}{2}\sum_\ell(1-\tau_\ell^x).
\end{aligned}}
\end{equation}
The first term is the energy per occupied flux plaquette.  The $m$ term is the
line energy of the Gauss-law-reconstructed boundary.  The $t$ term is the local
boundary-dressed Wilson transition, and the $\kappa$ term is a diagonal
boundary interaction.

Using Eq.~\eqref{eq:gauss-reduction}, the Hamiltonian reduces exactly to
\begin{equation}
{
\begin{aligned}
  H_\mathrm{phys}={}&
  \frac{\varepsilon_E}{2}\sum_p(1-\sigma_p^x)
  +\frac{m}{2}\sum_\ell
    \left(1-\sigma_{p_1(\ell)}^x\sigma_{p_2(\ell)}^x\right) \\
  &-\kappa\sum_p\prod_{p'\sim p}\sigma_{p'}^x
  -t\sum_p\bar\sigma_p^z.
\end{aligned}}
\label{eq:reduced-H}
\end{equation}
Here $\bar\sigma_p^z$ is the induced action of $W_p$ in the physical basis.
Equation~\eqref{eq:reduced-H} is the Ising-type plaquette Hamiltonian used in
all exact diagonalization calculations below.  The unreduced form Eq.~\eqref{eq:H} supplies the local
circuit representation and the Gauss-law syndromes; the reduced form supplies the
minimal exact physical Hilbert space.

\subsection{Top-form flux and the Brown--Teitelboim correspondence}
\label{sec:BT}

In $(2+1)$ dimensions, $H=dB$ is a top form.  The gauge dynamics has no
local gauge polarizations, and its source-free field strength is constant.
This matches the kinematic top-form sector of the Brown--Teitelboim setup in
$(3+1)$ dimensions, where a four-form field strength is changed by nucleation
of charged membrane worldvolumes~\cite{BrownTeitelboim1987,BrownTeitelboim1988}.

The Hamiltonian electric flux is
\[
  Q_E(t)=\frac{1}{2\pi}\int_{T^2}\iota_{\partial_t}H
        =\frac{1}{2\pi}\int_{T^2}E_{12}(\mathbf x,t)\,d^2x.
\]
The $\mathbb Z_2$ observable $M(t)$ instead counts plaquettes in one of two
flux states; it is an occupied-area observable, not a signed discretization of
$Q_E$.  The present model is therefore a lower-dimensional Hamiltonian analog
of top-form flux discharge.  It does not identify an equal-time flux region
with the timelike worldsheet of a charged string, and it does not include
gravity.

The local operator $\mathcal W_p$ changes the flux configuration by one plaquette while
preserving the induced boundary constraint.  Since $\mathcal W_p$ changes $M$ by one,
$M\bmod2$ is not a superselection label when $t\neq0$.  Fixed flux sectors arise
only when the charged boundary-changing term is absent or when an additional
global restriction is imposed.

\section{Quench protocol and observables}
\label{sec:hamiltonian}

\subsection{Paired-string initial state: tube and cap configurations}

The quench compares two gauge-invariant electric-flux fillings with the same
closed boundary.  The spatial topology is the two-torus $T^2$ (or a
cylinder $S^1\times\mathbb{R}$), with two non-contractible closed boundaries
$\gamma_1$ and $\gamma_2$ wrapping the compact $x$-direction at link rows
$y=y_1$ and $y=y_2$.  In the $\Z_2$ electric basis these boundaries are the
links with $\tau^x_\ell=-1$.  Gauss-law projection forces such activated links to
form closed one-cycles; open arcs are projected out of the physical sector.

Given the pair $\gamma_1\cup\gamma_2$, there are two natural electric-flux regions
with the same boundary:
\begin{itemize}
\item \textbf{Tube} configuration: $R_\mathrm{tube}$ is the plaquette strip
between $\gamma_1$ and $\gamma_2$, with area
$|R_\mathrm{tube}|=(y_2-y_1)L_x$.
\item \textbf{Cap} configuration: $R_\mathrm{cap}$ is the complementary strip
outside $\gamma_1,\gamma_2$ on the torus, with area
$|R_\mathrm{cap}|=(L_y-y_2+y_1)L_x$.
\end{itemize}
The two electric-basis product states
$\ket{\psi_0^{\rm tube}}=W[R_\mathrm{tube},\gamma_1\!\cup\!\gamma_2]\ket{+}^{\otimes N}$
and
$\ket{\psi_0^{\rm cap}}=W[R_\mathrm{cap},\gamma_1\!\cup\!\gamma_2]\ket{+}^{\otimes N}$
are both gauge invariant and have the same boundary links.  They differ by a
closed Wilson surface winding the full spatial torus.  The electric-tension
term assigns different energies to the two fillings whenever their areas
differ.

The electric-plus-mass parts of the initial energies are
\begin{align}
E_{0,\,E+m}^{\rm tube} &= \varepsilon_E\,(y_2-y_1)L_x
      +m\,|\gamma_1\cup\gamma_2|, \\
E_{0,\,E+m}^{\rm cap}  &= \varepsilon_E\,(L_y-y_2+y_1)L_x
      +m\,|\gamma_1\cup\gamma_2| .
\end{align}
The $\kappa$ contribution is a common boundary-dependent diagonal term for the
same pair $\gamma_1\cup\gamma_2$, and the $t$ contribution vanishes in these
initial electric-basis product states.  Thus the tube--cap energy difference is
purely electric:
\begin{equation}
E_0^{\rm cap}-E_0^{\rm tube}
=\varepsilon_E\bigl(|R_\mathrm{cap}|-|R_\mathrm{tube}|\bigr).
\end{equation}
For $y_2-y_1<L_y/2$ the tube is lower; for $y_2-y_1>L_y/2$ the cap is lower;
at $y_2-y_1=L_y/2$ they are degenerate.  This supplies the metastable
high-flux/low-flux pair of initial states whose differential dynamics resolves
the flux-discharge process.

\subsection{Preparation circuit}

Either initial state $\ket{\psi_0^\mathrm{tube}}$ or $\ket{\psi_0^\mathrm{cap}}$
is produced by a depth-2 single-qubit preparation circuit.  Starting from the
hardware computational state $|0\rangle^{\otimes N_q}$, apply $X$ gates to the
plaquette and link qubits that should be flipped in the electric basis, and
then apply a Hadamard layer to all qubits.  Equivalently, after the Hadamard
layer prepares the electric-basis vacuum, the same excitation pattern is
implemented by $Z$ gates on every plaquette qubit $p\in R_\mathrm{tube}$
(respectively $R_\mathrm{cap}$) and every link qubit
$\ell\in\gamma_1\cup\gamma_2$.  The preparation uses no entangling gates.

\subsection{Quench and Trotter decomposition}

The quench evolves $\ket{\psi_0}$ under $H$ of Eq.~(\ref{eq:H}) for total
time $T$ using a second-order Strang-Trotter decomposition,
\begin{equation}
  e^{-i H \delta t} \approx e^{-i H_A \delta t/2} e^{-i H_B \delta t} e^{-i H_A \delta t/2},
\end{equation}
with $H_A$ the single-qubit electric and mass terms ($\sigma^x_p$ and $\tau^x_\ell$),
and $H_B$ the multi-qubit Wilson-surface and link-electric plaquette terms. Each
Pauli-string exponential in $H_B$ acts on five qubits (Wilson-surface:
$\sigma^z_p$ plus four $\tau^z$) or four qubits (the $H_\kappa$ Pauli string
$\tau^x \tau^x \tau^x \tau^x$); both are standard diagonal-in-a-basis
multi-qubit rotations implementable by CNOT staircase $+$ single-qubit
rotation $+$ CNOT staircase.

\subsection{Observables}

The quantities most directly probing flux-domain discharge dynamics are local and
easy to measure in the $X$-basis:
\begin{align}
  M(t) &= \sum_p \left\langle \frac{1 - \sigma^x_p}{2} \right\rangle_t
        \quad (\text{occupied-flux area}) \label{eq:Mobs}\\
  L(t) &= \sum_\ell \left\langle \frac{1 - \tau^x_\ell}{2} \right\rangle_t
        \quad (\text{boundary length}). \label{eq:Lobs}
\end{align}
Both are sums of single-qubit $X$ expectations, requiring only a basis-rotation
layer before readout. Initially $M(0) = |R|$ and $L(0) = |\gamma|$.

Richer observables include spatial distributions
$\mu_p(t) = \ev{(1-\sigma^x_p)/2}_t$ and $\lambda_\ell(t)$ and
connected-component counts of the excited-plaquette set (``flux-domain multiplicity''), all obtained from the same electric-basis string bits.
Entanglement diagnostics require additional randomized measurements; classical
shadow methods can access selected reduced density matrices with the usual
subsystem-dependent overhead \cite{HuangShadows}.

\section{Hardware embedding}
\label{sec:hardware}

\begin{figure}[!htbp]
\centering
\resizebox{0.98\textwidth}{!}{%
\begin{tikzpicture}[scale=1.0, every node/.style={font=\footnotesize},
                    box/.style={draw, thick, rounded corners, minimum height=1.1cm,
                                minimum width=2.4cm, align=center, fill=blue!5},
                    boxp/.style={draw, thick, rounded corners, minimum height=1.1cm,
                                 minimum width=2.2cm, align=center, fill=red!10},
                    boxm/.style={draw, thick, rounded corners, minimum height=1.1cm,
                                 minimum width=2.3cm, align=center, fill=green!10}]
\node[box] (init) at (0, 0) {$|0\rangle^{\otimes N_q}$\\\scriptsize $N_q = L^2 + N_\ell$};
\node[boxp] (prep) at (3.1, 0) {State prep\\\scriptsize $X$-flips + $H$ layer\\\scriptsize depth 2};
\node[box] (trotN) at (6.5, 0) {Trotter step $\times\,N_T$\\\scriptsize $U_\mathrm{Trot}(\delta t)$\\\scriptsize depth $32$--$128$ each};
\node[boxm] (meas) at (10.1, 0) {Measurement\\\scriptsize $H$-layer $+$ $Z$-readout\\\scriptsize depth 1};
\node[draw, thick, dashed, rounded corners, minimum height=1.1cm, minimum width=2.6cm, align=center,
      fill=yellow!20] (post) at (13.5, 0) {Gauss post-sel\\\scriptsize $G_\ell(b)=+1\ \forall \ell$};
\draw[->, thick] (init) -- (prep);
\draw[->, thick] (prep) -- (trotN);
\draw[->, thick] (trotN) -- (meas);
\draw[->, thick] (meas) -- (post);
% output arrow
%\draw[->, thick] (post) -- ++(1.5, 0) node[right] {$\{M, L, N_\mathrm{tube}, N_\mathrm{cap}\}$};
\draw[->, thick] (post) -- ++(1.9, 0) node[right] {$\begin{matrix}
  M \\
  L \\
  N_{\mathrm{tube}} \\
  N_{\mathrm{cap}}
\end{matrix}$};

% Expand the Trotter block beneath
\node[draw, thick, rounded corners, minimum width=1.1cm, minimum height=0.6cm, fill=blue!20] (e1) at (2.2, -2.0) {\scriptsize $e^{-i\frac{\delta t}{2} H_E}$};
\node[draw, thick, rounded corners, minimum width=1.1cm, minimum height=0.6cm, fill=blue!20] (m1) at (4.0, -2.0) {\scriptsize $e^{-i\frac{\delta t}{2} H_m}$};
\node[draw, thick, rounded corners, minimum width=1.1cm, minimum height=0.6cm, fill=green!25] (kappa) at (5.8, -2.0) {\scriptsize $e^{-i\delta t H_\kappa}$};
\node[draw, thick, rounded corners, minimum width=1.1cm, minimum height=0.6cm, fill=red!25] (t) at (7.6, -2.0) {\scriptsize $e^{-i\delta t H_t}$};
\node[draw, thick, rounded corners, minimum width=1.1cm, minimum height=0.6cm, fill=blue!20] (m2) at (9.4, -2.0) {\scriptsize $e^{-i\frac{\delta t}{2} H_m}$};
\node[draw, thick, rounded corners, minimum width=1.1cm, minimum height=0.6cm, fill=blue!20] (e2) at (11.2, -2.0) {\scriptsize $e^{-i\frac{\delta t}{2} H_E}$};
\draw[->] (e1) -- (m1);
\draw[->] (m1) -- (kappa);
\draw[->] (kappa) -- (t);
\draw[->] (t) -- (m2);
\draw[->] (m2) -- (e2);
\draw[dotted, thick] (trotN.south west) -- (e1.north west);
\draw[dotted, thick] (trotN.south east) -- (e2.north east);

% annotation below Trotter blocks
\node[align=center] at (2.2, -2.7) {\tiny diagonal\\\tiny single-qubit};
\node[align=center] at (4.0, -2.7) {\tiny diagonal\\\tiny single-qubit};
\node[align=center] at (5.8, -2.7) {\tiny 4-body\\\tiny $6$-$24\,L^2$ CNOTs};
\node[align=center] at (7.6, -2.7) {\tiny 5-body\\\tiny $8\,L^2$ CNOTs};
\node[align=center] at (9.4, -2.7) {\tiny diagonal\\\tiny single-qubit};
\node[align=center] at (11.2, -2.7) {\tiny diagonal\\\tiny single-qubit};
\end{tikzpicture}%
}
\caption{End-to-end digital quantum simulation pipeline. The $N_q =
L^2 + N_\ell$ physical qubits are initialized in $|0\rangle^{\otimes
N_q}$, prepared in the tube or cap initial state by a depth-2
single-qubit circuit ($X$-flips on a chosen plaquette sublattice
followed by Hadamards on all qubits), evolved by $N_T$ second-order
symmetric Strang--Trotter steps (expanded lower row: each step
interleaves the single-qubit electric terms with the entangling
terms $H_\kappa$ (4-body link-electric plaquette interaction) and $H_t$
(5-body Wilson-surface)), and finally measured in the $X$-basis via a
Hadamard layer followed by $Z$-readout.  The displayed $H_\kappa$ cost is
$6L^2$ CNOTs when the four boundary-link qubits form a native path and
$24L^2$ CNOTs on a minimal plaquette-link star using routed dirty-ancilla
parity accumulation.  Gauss-law post-selection on all link constraints
$\mathcal G_\ell(b)=+1$ rejects gauge-violating bitstrings.}
\label{fig:pipeline}
\end{figure}
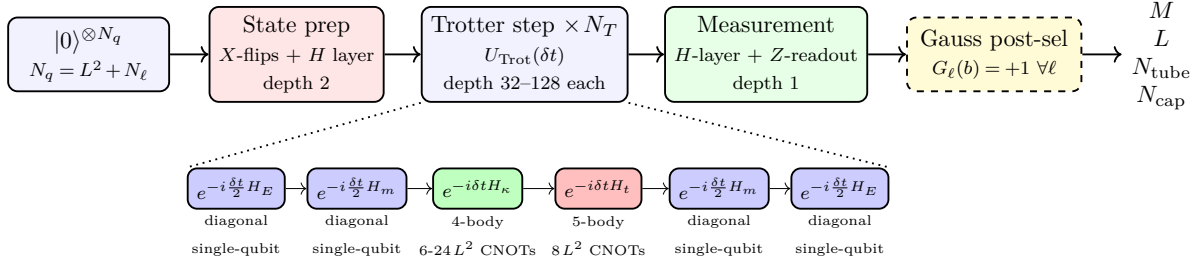

The model has two qubit species--plaquette (gauge) and link (matter)--with a
bipartite incidence structure.  Each plaquette qubit couples to its four
boundary link qubits through the Wilson-surface term and through the Gauss
generators; each interior link qubit touches two plaquettes.  The logical graph
$G_{\rm log}$ has one edge between every plaquette and every link on its
boundary.

Boundary conditions affect both qubit count and routing.  The exact diagonalization benchmarks in
Sec.~\ref{sec:ED} use a torus, for which $N_\ell=2L^2$ and
$N_q=3L^2$.  A planar open patch has $N_\ell=2L(L+1)$ and
$N_q=3L^2+2L$.  A cylinder periodic in the string direction has
$N_\ell=L(2L+1)$ and $N_q=3L^2+L$.  A planar device can realize the torus
Hamiltonian only with boundary wrap-around routing; without that routing it
implements the open-patch or cylindrical variant.

\begin{table}[!htbp]
\centering
\begin{tabular}{@{}rrrrr@{}}
\toprule
$L$ & $N_p$ & $N_q$ torus & $N_q$ cylinder & $N_q$ open patch \\
\midrule
3 & 9  & 27  & 30  & 33 \\
4 & 16 & 48  & 52  & 56 \\
5 & 25 & 75  & 80  & 85 \\
6 & 36 & 108 & 114 & 120 \\
\bottomrule
\end{tabular}
\caption{Data-qubit counts for an $L\times L$ plaquette lattice.  The torus
count is the one corresponding to the exact diagonalization topology.  The open-patch count is the
one naturally embedded on a planar nearest-neighbor device without periodic
wrap-around routing.}
\label{tab:qubit-count}
\end{table}

\subsection{Square-lattice and IBM heavy-hex embeddings}

A rotated-square device graph gives a native realization of the plaquette-link
star: place plaquette qubits on one sublattice and the horizontal/vertical link
qubits on neighboring sites.  In this layout every plaquette qubit is adjacent
to the four link qubits in its boundary.  The 5-body Wilson-surface rotation is
therefore native on the logical star.  Periodic boundary conditions require
additional routing along the outer boundary of the planar chip; the open-patch
and cylindrical variants avoid one or both wrap-around routes.

The 4-body $H_\kappa$ term is more restrictive.  It acts only on the four link
qubits around a plaquette.  In the minimal plaquette-link star those four link
qubits are not mutually adjacent, so the six-CNOT linear staircase is native
only on an augmented layout where the four boundary links form a hardware path.
On the minimal star, $H_\kappa$ must be compiled through the central plaquette
qubit as a dirty routing ancilla, or implemented after SWAP routing.  This is
the main correction to the hardware resource estimate.

IBM's heavy-hex connectivity has maximum degree three, while a bulk plaquette star has
degree four.  A heavy-hex implementation therefore routes at least one edge of
each bulk plaquette star and also routes the $H_\kappa$ four-link parity unless
additional ancillae are allocated.  The heavy-hex numbers are consequently
compiler-dependent.  The resource table below gives logical counts for three
schedules; platform-specific compilation should be performed after choosing the
boundary condition and device subgraph.

\subsection{Elementary gate sequences for the four Hamiltonian terms}
\label{sec:gates}

The rotation convention is
\[
  R_\alpha(\theta)=\exp(-i\theta P_\alpha/2),
\]
with $P_\alpha=X,Y,Z$ as appropriate.  With this convention the nontrivial part
of $\exp[-i\delta t\,\frac{\varepsilon_E}{2}(1-X)]$ is implemented by
$R_x(-\varepsilon_E\delta t)$; similarly the matter-mass term uses
$R_x(-m\delta t)$.

\paragraph{$H_E$ and $H_m$ single-qubit rotations.}
The terms $H_E=\frac{\varepsilon_E}{2}\sum_p(1-\sigma_p^x)$ and
$H_m=\frac{m}{2}\sum_\ell(1-\tau_\ell^x)$ are parallel layers of
single-qubit $X$ rotations, up to global phases:
\[
  U_E(\delta t)=\prod_p R_x^{(p)}(-\varepsilon_E\delta t),\qquad
  U_m(\delta t)=\prod_\ell R_x^{(\ell)}(-m\delta t).
\]

\paragraph{$H_\kappa$ path-native circuit.}
For a plaquette whose four boundary-link qubits form a hardware path
$\ell_1-\ell_2-\ell_3-\ell_4$, the interaction
$-\kappa\prod_{\ell\in\partial p}\tau_\ell^x$ is implemented by Hadamards on
the four links, a four-qubit $Z$-parity rotation, and the inverse Hadamards:
\begin{quote}
\centering
\small\sf\noindent
$H_{\ell_1}H_{\ell_2}H_{\ell_3}H_{\ell_4}$\\
$\textsc{cnot}(\ell_1,\ell_2);\ \textsc{cnot}(\ell_2,\ell_3);\ \textsc{cnot}(\ell_3,\ell_4)$\\
$R_z^{(\ell_4)}(-2\kappa\delta t)$\\
$\textsc{cnot}(\ell_3,\ell_4);\ \textsc{cnot}(\ell_2,\ell_3);\ \textsc{cnot}(\ell_1,\ell_2)$\\
$H_{\ell_1}H_{\ell_2}H_{\ell_3}H_{\ell_4}$ .
\end{quote}
The sign of the $R_z$ angle follows from
$e^{-i\delta t(-\kappa P)}=e^{+i\kappa\delta t P}$.  This path-native version
uses 6 CNOTs and 6 CNOT layers per plaquette.

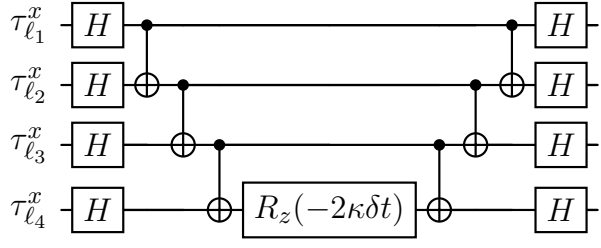
\begin{figure}[!htbp]
\centering
\resizebox{0.50\linewidth}{!}{%
\begin{quantikz}[column sep=0.14cm, row sep=0.2cm]
\lstick{$\tau^x_{\ell_1}$} & \gate{H} & \ctrl{1} & \qw      & \qw      & \qw      & \qw      & \qw      & \ctrl{1} & \gate{H} & \qw \\
\lstick{$\tau^x_{\ell_2}$} & \gate{H} & \targ{}  & \ctrl{1} & \qw      & \qw      & \qw      & \ctrl{1} & \targ{}  & \gate{H} & \qw \\
\lstick{$\tau^x_{\ell_3}$} & \gate{H} & \qw      & \targ{}  & \ctrl{1} & \qw      & \ctrl{1} & \targ{}  & \qw      & \gate{H} & \qw \\
\lstick{$\tau^x_{\ell_4}$} & \gate{H} & \qw      & \qw      & \targ{}  & \gate{R_z(-2\kappa\delta t)} & \targ{}  & \qw      & \qw      & \gate{H} & \qw
\end{quantikz}%
}
\caption{Path-native circuit for the 4-body link-electric plaquette rotation
$\exp(i\kappa\delta t\prod_{\ell\in\partial p}\tau_\ell^x)$.  The circuit is
native only when the four boundary-link qubits form a connected hardware path.
On a minimal plaquette-link star, the same logical rotation requires routing.}
\label{fig:circuit_matter}
\end{figure}

\paragraph{$H_\kappa$ routed star circuit.}
On a minimal star, an effective CNOT between two link qubits $a$ and $b$ routed
through the central plaquette qubit $c$ can be implemented with $c$ as a dirty
ancilla by
\[
  \textsc{cnot}(c,b)\,\textsc{cnot}(a,c)\,\textsc{cnot}(c,b)\,\textsc{cnot}(a,c),
\]
which restores $c$ and realizes $\textsc{cnot}(a,b)$.  Replacing each of the
three CNOTs in the parity-compute half of the four-link staircase by this
routed CNOT, and doing the same in the uncompute half, gives 24 CNOTs per
plaquette for $H_\kappa$.  This count is conservative and assumes no additional
clean ancillae.

\paragraph{$H_t$ star-native Wilson-surface circuit.}
The Wilson-surface term
\[
  -t\,\sigma^z_p\prod_{\ell\in\partial p}\tau^z_\ell
\]
acts on the central plaquette qubit and its four boundary-link qubits.  On the
plaquette-link star, its parity can be accumulated directly onto the plaquette
qubit:
\begin{quote}
\centering
\small\sf\noindent
$\textsc{cnot}(\ell_1,p);\ \textsc{cnot}(\ell_2,p);\ \textsc{cnot}(\ell_3,p);\ \textsc{cnot}(\ell_4,p)$\\
$R_z^{(p)}(-2t\delta t)$\\
$\textsc{cnot}(\ell_4,p);\ \textsc{cnot}(\ell_3,p);\ \textsc{cnot}(\ell_2,p);\ \textsc{cnot}(\ell_1,p)$ .
\end{quote}
This uses 8 CNOTs and 8 CNOT layers per plaquette.  CNOT direction is a
hardware-convention issue; reversing a directed edge costs only local basis
changes.

\begin{figure}[!htbp]
\centering
\resizebox{0.5\linewidth}{!}{%
\begin{quantikz}[column sep=0.14cm, row sep=0.4cm]
\lstick{$\sigma^z_p$}      & \targ{}  & \targ{}  & \targ{}  & \targ{}  & \gate{R_z(-2t\delta t)} & \targ{}  & \targ{}  & \targ{}  & \targ{}  & \qw \\
\lstick{$\tau^z_{\ell_1}$} & \ctrl{-1} & \qw      & \qw      & \qw      & \qw                         & \qw      & \qw      & \qw      & \ctrl{-1} & \qw \\
\lstick{$\tau^z_{\ell_2}$} & \qw      & \ctrl{-2} & \qw      & \qw      & \qw                         & \qw      & \qw      & \ctrl{-2} & \qw      & \qw \\
\lstick{$\tau^z_{\ell_3}$} & \qw      & \qw      & \ctrl{-3} & \qw      & \qw                         & \qw      & \ctrl{-3} & \qw      & \qw      & \qw \\
\lstick{$\tau^z_{\ell_4}$} & \qw      & \qw      & \qw      & \ctrl{-4} & \qw                         & \ctrl{-4} & \qw      & \qw      & \qw      & \qw
\end{quantikz}%
}
\caption{Star-native circuit for the 5-body Wilson-surface rotation
$\exp(i t\delta t\,\sigma^z_p\prod_{\ell\in\partial p}\tau^z_\ell)$.  The four
boundary links control CNOTs into the plaquette accumulator, a single
$R_z(-2t\delta t)$ applies the parity phase, and the CNOTs are reversed.  The
operator flips one plaquette electric flux and the four adjacent link
occupations in the electric basis, preserving all four affected Gauss-laws.}
\label{fig:circuit_wilson}
\end{figure}
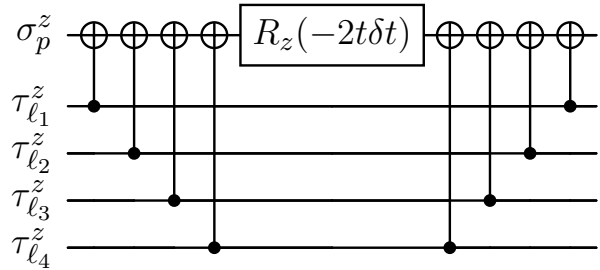

\paragraph{4-coloring parallelization.}
Open square patches and even-$L$ square tori admit a 4-coloring such that no
two plaquettes of the same color share a boundary link.  Odd-$L$ tori require a
modified coloring and have a slightly larger constant depth.  Within one color
class, all $H_t$ circuits act on disjoint qubit sets and run in parallel.  The
same is true for $H_\kappa$ after choosing either a path-native or routed-star
implementation.  The resulting logical CNOT depth per Strang-Trotter step is
\[
  D_1=32\quad (H_t\ \text{only}),\qquad
  D_1=56\quad (H_t+H_\kappa\ \text{with path-native }H_\kappa),
\]
\[
  D_1\simeq128\quad (H_t+H_\kappa\ \text{on a minimal routed star}).
\]
For open patches and even-$L$ tori, all three depths are independent of $L$
before periodic-boundary routing.

\begin{figure}[!htbp]
\centering
\begin{tikzpicture}[scale=0.6, every node/.style={font=\scriptsize}]
\foreach \x in {0,...,3} {
  \foreach \y in {0,...,3} {
    \pgfmathsetmacro{\cc}{mod(\x,2) + 2*mod(\y,2)}
    \pgfmathtruncatemacro{\cci}{\cc}
    \ifcase\cci%
       \def\col{red!30}\def\lab{1}%
    \or \def\col{blue!30}\def\lab{2}%
    \or \def\col{green!40}\def\lab{3}%
    \or \def\col{orange!40}\def\lab{4}%
    \fi
    \fill[\col] (\x,\y) rectangle (\x+1, \y+1);
    \node at (\x+0.5, \y+0.5) {\lab};
  }
}
\foreach \x in {0,...,4} \draw[thick, gray] (\x,0) -- (\x,4);
\foreach \y in {0,...,4} \draw[thick, gray] (0,\y) -- (4,\y);
\foreach \x in {0,...,3} \foreach \y in {0,...,3}
  \draw[fill=white, thick] (\x+0.5, \y+0.5) circle (0.14);
\node[right] at (4.5, 3.2) {\small Colors $=$ time slots};
\node[right] at (4.5, 2.6) {\small 1--4 act sequentially};
\node[right] at (4.5, 2.0) {\small same-color plaquettes};
\node[right] at (4.5, 1.4) {\small run in parallel};
\node[right] at (4.5, 0.7) {\small $D_1=32,56,$ or $128$};
\end{tikzpicture}
\caption{Four-coloring used to parallelize plaquette terms.  The coloring
keeps same-color plaquettes from sharing boundary links.  The depth per step is
constant in lattice size but depends on the $H_\kappa$ connectivity: 56 layers
for path-native four-link parity rotations and about 128 layers on the minimal
routed star.}
\label{fig:coloring}
\end{figure}
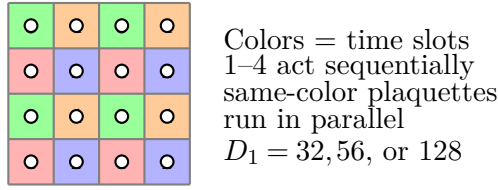

\subsection{Trotter decomposition and error analysis}
\label{sec:trotter}

The second-order symmetric Strang--Trotter splitting is
\begin{equation}
\begin{aligned}
  U_{\rm Trot}(\delta t)=\;& e^{-i\frac{\delta t}{2}H_E}
   e^{-i\frac{\delta t}{2}H_m}
   e^{-i\delta t H_\kappa} \\
  &\times e^{-i\delta t H_t}
   e^{-i\frac{\delta t}{2}H_m}e^{-i\frac{\delta t}{2}H_E} .
\end{aligned}
\label{eq:trotter}
\end{equation}
The local error is $\mathcal{O}(\delta t^3)$ and is controlled by nested
commutators among the four Pauli-string groups.

The useful elementary commutators are as follows.  Let
$A_p=\sigma^z_p\prod_{\ell\in\partial p}\tau^z_\ell$ and
$K_q=\prod_{\ell\in\partial q}\tau^x_\ell$.  The same-plaquette commutator
vanishes,
\begin{equation}
  [A_p,K_p]=0,
\end{equation}
because the two strings overlap on four link qubits.  The nonzero
$H_t$--$H_\kappa$ commutators come from adjacent plaquettes.  If $p$ and $q$
share the link $s$, then, up to the sign set by orientation conventions,
\begin{equation}
  [H_t^{(p)},H_\kappa^{(q)}]
  =2i t\kappa\,\sigma^z_p\tau^y_s
  \prod_{\ell\in\partial p\setminus s}\tau^z_\ell
  \prod_{\ell\in\partial q\setminus s}\tau^x_\ell .
\end{equation}
The single-qubit commutators are
\begin{align}
  [H_t^{(p)},H_E^{(p)}]
  &= i t\varepsilon_E\,\sigma^y_p\prod_{\ell\in\partial p}\tau^z_\ell, \\
  [H_t^{(p)},H_m^{(\ell)}]
  &= i t m\,\sigma^z_p\tau^y_\ell
     \prod_{\ell'\in\partial p\setminus\{\ell\}}\tau^z_{\ell'} .
\end{align}
The signs in these formulas are immaterial for the norm estimates, but the
support is the operative point: $[H_t,H_\kappa]$ is an adjacent-plaquette
commutator rather than a same-plaquette commutator.

On an $L\times L$ torus these supports give the conservative bounds
\begin{align}
  \|[H_t,H_E]\| &\le t\varepsilon_E L^2, \\
  \|[H_t,H_m]\| &\le 4tmL^2, \\
  \|[H_t,H_\kappa]\| &\le 8t\kappa L^2,
\end{align}
where the last line counts the four adjacent plaquettes for each plaquette and
the factor of two from one-link anticommutation.

\paragraph{Numerical commutator norms on the $L=4$ lattice.}
The operator norms on the $4\times4$
Gauss-law-projected Hilbert space (dimension $2^{16}$) with
$\varepsilon_E=m=1$ and $t=\kappa=0.4$:
\begin{equation}
  \begin{aligned}
    \|[H_t,H_E]\|_{\rm op} &= 6.40,
    & t\varepsilon_E L^2 &= 6.40, \\
    \|[H_t,H_\kappa]\|_{\rm op} &= 10.24,
    & 8t\kappa L^2 &= 20.48, \\
    \|[H_t,H_m]\|_{\rm op} &= 14.40,
    & 4tmL^2 &= 25.60 .
  \end{aligned}
\end{equation}
The corrected support counting places all three norms below the simple
operator-sum bounds.  The corresponding nested commutators used in the
observable-level Trotter check are
$\|[H_t,[H_t,H_E]]\|_{\rm op}=5.12$,
$\|[H_t,[H_t,H_\kappa]]\|_{\rm op}=32.77$, and
$\|[H_t,[H_t,H_m]]\|_{\rm op}=23.04$.

\paragraph{Observable-level Trotter error: direct measurement.}
The formal operator-norm bound
\begin{equation}
  \epsilon_{\rm Trot}(T) \lesssim \frac{T\delta t^2}{12}
  \sum_{A,B,C}\|[H_A,[H_B,H_C]]\|
\end{equation}
is intentionally conservative.  The observable error is measured
directly by running the Strang--Trotter integrator on the $4\times4$ lattice at
$\delta t\in\{0.05,0.1,0.25,0.5\}$ and comparing against the Krylov reference
at the same lattice. The maximum errors over $t\in[0,10]$ are
\begin{equation}
  \begin{array}{c|cccc}
    \delta t & 0.05 & 0.10 & 0.25 & 0.50 \\
    \hline
    \max|\Delta M_{\rm cap}| & 0.003 & 0.002 & 0.016 & 0.073 \\
    \max|\Delta L_{\rm cap}| & 0.013 & 0.008 & 0.056 & 0.231
  \end{array}
\end{equation}
The error scales as $\mathcal{O}(\delta t^2)$, as expected for a second-order
Strang-Trotter scheme.  At $\delta t=0.5$ the maximum $M_{\rm cap}$ error is $0.073$,
about $0.6\%$ of the typical signal magnitude $\langle M_{\rm cap}\rangle\sim
8$--$12$.  This lies below the target observable-bias scale used in
Sec.~\ref{sec:error}.  Future parameter sweeps can reuse the same observable-level
Trotter check with the corresponding Hamiltonian normalization.

\subsection{State-preparation circuit for tube and cap initial states}
\label{sec:prep}

The tube and cap initial states defined in Sec.~\ref{sec:hamiltonian} are
classical computational-basis states in the electric basis (every $\sigma^x_p$
and $\tau^x_\ell$ is $\pm 1$). The preparation circuit is therefore very
shallow.

Let $\ket{0_\sigma}$ denote the $\sigma^x = +1$ state of a plaquette qubit
and $\ket{0_\tau}$ the $\tau^x = +1$ state of a link qubit. The
hardware-native ``all-zero'' state $\ket{0}^{\otimes N}$ is the $\sigma^z =
+1, \tau^z = +1$ state, whereas the target preparation begins in the electric-basis vacuum.

The preparation circuit acts in two layers:
\begin{enumerate}
\item \textbf{Hadamard layer:} apply $H$ to every plaquette and every link
qubit. After this layer, every qubit is in the $\sigma^x=+1$ (resp.
$\tau^x=+1$) eigenstate. This is the electric-basis vacuum.
\item \textbf{Tube/cap excitation layer:} for the tube preparation, apply
single-qubit $Z$-flips to the $\sigma^x$-basis on every plaquette in the
tube-strip $R_{\rm tube}$ and on every link in the boundary closed strings
$\gamma_1, \gamma_2$. Equivalently in the computational basis, apply $X$
gates on each of those qubits before the Hadamard layer (since $H X H = Z$).
For the cap preparation, replace $R_{\rm tube}$ by $R_{\rm cap}$.
\end{enumerate}

\noindent The total preparation circuit is therefore depth 2 (one $X$-layer
plus one $H$-layer), uses no entangling gates, and is identical for tube and
cap preparations except for the choice of which plaquettes receive the
$X$-flip. This makes the tube-vs-cap comparison controlled on hardware: the two runs
differ only by the classical bit-flip pattern at $t=0$.

\paragraph{Worked $4\times 4$ example.}
For the $L_x=L_y=4$ torus with closed strings at link-rows $y_1=0, y_2=1$
(separation 1), the tube region $R_{\rm tube}$ is the single plaquette
row $y=0$ (4 plaquettes) and the boundary strings
$\gamma_1\cup\gamma_2$ are the two horizontal link-rows at $y=0$ and
$y=1$ (8 links). The tube initial-state preparation circuit is:

\begin{quote}\small\sf\noindent
Initialize 48 torus data qubits in $|0\rangle^{\otimes 48}$\\
$X^{(p)}$ for each plaquette $p$ in row $y=0$ (4 X-gates)\\
$X^{(\ell)}$ for each link $\ell$ in rows $y=0$ or $y=1$ (8 X-gates)\\
$H^{(p)}$ for every plaquette qubit (16 H-gates)\\
$H^{(\ell)}$ for every torus link qubit (32 H-gates)
\end{quote}
\noindent Total: 12 $X$-gates and 48 $H$-gates, distributed in 2 layers
of single-qubit operations on the torus data register. The cap initial state
replaces the 4 plaquette $X$-gates by 12 ($\sigma^x=-1$ on rows $y=1,2,3$
instead of $y=0$), while keeping the 8 link $X$-gates identical.  For an
independent single-qubit error rate $p_1\sim10^{-4}$, the preparation-only
success factor is $(1-p_1)^{60}\approx0.994$ before Trotter evolution.  The
open-patch realization uses the 56-qubit register listed in
Table~\ref{tab:qubit-count}.

\subsection{Measurement circuit and Gauss-law post-selection}
\label{sec:meas}

All observables of interest---total electric occupied-flux area $M(t)$, total
boundary length $L(t)$, and tube-vs-cap region occupations
$N_\mathrm{tube}(t), N_\mathrm{cap}(t)$---are diagonal in the electric basis.
Measurement therefore consists of:
\begin{enumerate}
\item A global Hadamard layer on all plaquette and link qubits, mapping
$\sigma^x \to \sigma^z$ and $\tau^x \to \tau^z$.
\item Computational-basis readout of all qubits.
\end{enumerate}
The readout produces a bitstring $b \in \{0,1\}^N$, from which one computes
$M, L, N_\mathrm{tube}, N_\mathrm{cap}$ by classical post-processing.

\paragraph{Gauss-law post-selection.}
The 1-form Gauss-law ${\mathcal G}_\ell = \tau^x_\ell \prod_{p \ni \ell} \sigma^x_p
= +1$ is a constraint that physical states must satisfy on every link.
Hardware noise---bit-flips, dephasing, leakage---generically takes the
state out of the Gauss-law-constrained subspace. After computational-basis
readout, the bitstring $b$ contains both the plaquette gauge bits
$\{s_p\}$ and the link matter bits $\{t_\ell\}$, and the constraint is checked directly as
\begin{equation}
  \mathcal G_\ell(b) = (-1)^{t_\ell + \sum_{p\ni\ell} s_p}
\end{equation}
for every link. Shots are discarded whenever $\mathcal G_\ell(b) = -1$ on any
link. This gauge-aware post-selection is the detection-only implementation of the Gauss-law coding principle developed in Refs.~\cite{Wiebe2023,Wiebe2026} and is operationally analogous to the 
symmetry-protection protocol of 
Ref.~\cite{Nguyen2022Schwinger}. In the present circuit, no active syndrome recovery is performed. The number of constraints
($N_\ell = 2L(L+1)$ on an $L\times L$ open lattice or $2L^2$ on the
torus) is much larger than the number of plaquette qubits, so
post-selection is statistically significant and must be included in the shot
budget.  It removes readout-frame and circuit errors that violate the link
Gauss-law; errors that remain inside the physical sector require the additional
mitigation steps discussed in Sec.~\ref{sec:error}.

\paragraph{Acceptance-rate and shot budget.}
Under an effective independent bit error rate $\bar p$ in the final
electric-basis readout frame, post-selection accepts error patterns that
satisfy the same Gaus-law constraints as the ideal bitstrings.  At small
$\bar p$, the leading rejection probability is therefore the probability of a
single data-bit error.  On an $L\times L$ torus, $N_q=3L^2$, so
\begin{equation}
  p_{\rm accept}^{\rm torus}(L,\bar p)
  =1-3L^2\bar p+\mathcal{O}(\bar p^2)
  \simeq \exp(-3L^2\bar p).
\end{equation}
For an open patch, $N_q=3L^2+2L$, giving
\begin{equation}
  p_{\rm accept}^{\rm OBC}(L,\bar p)
  =1-(3L^2+2L)\bar p+\mathcal{O}(\bar p^2).
\end{equation}
The estimate counts accepted correlated error strings only at higher order in
$\bar p$.  For $L=4$ and $\bar p\sim10^{-3}$, the torus acceptance is about
$0.95$ at leading order; for $L=6$ it is about $0.90$.  To achieve shot-noise
error $\epsilon_{\rm stat}$ on $M(t)/N_p$, one needs
$N_{\rm shots}\sim1/(\epsilon_{\rm stat}^2p_{\rm accept})$ raw shots up to the
state-dependent variance factor.  For $\epsilon_{\rm stat}=0.02$, this remains
a few thousand accepted shots per time point.

\paragraph{Worked $4\times 4$ readout circuit.}
The complete readout after the final Trotter step is:

\begin{quote}\small\sf\noindent
$H^{(q)}$ on all 48 torus data qubits (or 56 open-patch qubits)\\
Z-basis measurement on all data qubits\\
Classical processing: extract 16 plaquette bits $\{s_p\}$ and 32 torus link\\
\quad bits $\{t_\ell\}$; evaluate the 32 torus link Gauss constraints\\
\quad $G_\ell(b)$; discard shot if any $G_\ell(b)=-1$.\\
\quad From the accepted bitstring compute\\
\quad $M(b)=\sum_p s_p$, $L(b) = \sum_\ell t_\ell$,\\
\quad $N_{\rm tube}(b) = \sum_{p\in R_{\rm tube}} s_p$,\\
\quad $N_{\rm cap}(b) = \sum_{p\in R_{\rm cap}} s_p$.
\end{quote}

\paragraph{Cost.}
The measurement layer is depth-1 with no entangling gates. The total
measurement overhead is therefore one Hadamard layer plus one shot per
accepted bitstring, in addition to the post-selection bookkeeping; the
quantum cost is negligible compared to the Trotter circuit.

\subsection{Gate-count and depth analysis}

The resource count separates three logical schedules.  The first implements only
the Wilson-surface core $H_t$ together with the single-qubit electric and mass
terms.  The second implements the full $H_t+H_\kappa$ model assuming that each
plaquette's four boundary-link qubits form a native path.  The third implements
the full model on the minimal plaquette-link star, routing the $H_\kappa$ parity
through the central plaquette qubit as a dirty ancilla.

\begin{table}[!htbp]
\centering
\resizebox{\textwidth}{!}{%
\begin{tabular}{@{}lrrrr@{}}
\toprule
Schedule & CNOTs / plaquette & CNOTs / step & CNOT layers / step & Depth for $N_T=20$ \\
\midrule
Wilson core: $H_t$ only & $8$  & $8L^2$  & $32$  & $640$ \\
Full, path-native $H_\kappa$ & $14$ & $14L^2$ & $56$  & $1120$ \\
Full, routed-star $H_\kappa$ & $32$ & $32L^2$ & $128$ & $2560$ \\
\bottomrule
\end{tabular}
}
\caption{Logical entangling-gate resources per second-order Strang--Trotter step before
periodic-boundary routing.  The six-CNOT count for $H_\kappa$ is valid only
when the four boundary-link qubits form a native path.  On the minimal
plaquette-link star, $H_\kappa$ is a routed four-link parity rotation and costs
about 24 CNOTs per plaquette, giving the full-model routed-star count shown in
the last row.}
\label{tab:depth}
\end{table}

For an $L=4$ torus, the data-qubit count is 48; an open $4\times4$ patch uses
56 data qubits.  The CNOT counts per Trotter step are 128, 224, and 512 for the
three schedules in Table~\ref{tab:depth}.  Heavy-hex and planar torus
implementations add routing overhead that must be compiled on the chosen device
subgraph.  The table is therefore a logical schedule, not a device-calibrated
claim.

\subsection{Computational complexity analysis}
\label{sec:complexity}

The key resource scalings are parameterized by the linear lattice size $L$, the
evolution time $T$, the Trotter step $\delta t$, and the target observable
accuracy $\epsilon$.  Boundary conditions set the data-qubit count:
$N_q=3L^2$ on a torus, $N_q=3L^2+L$ on a cylinder, and
$N_q=3L^2+2L$ on an open square patch.

\paragraph{Entangling gates per Trotter step.}
The three logical schedules have
\begin{equation}
  C_1(L)=c_C L^2,
  \qquad
  c_C\in\{8,14,32\},
\end{equation}
for Wilson-core, path-native full, and routed-star full compilation,
respectively.  Platform routing for periodic boundaries or heavy-hex degree
constraints multiplies this by a device- and layout-dependent constant.

\paragraph{Trotter-step depth.}
For open patches and even-$L$ tori, four-coloring gives
\begin{equation}
  D_1\in\{32,56,128\}
\end{equation}
CNOT layers per step for the same three schedules, again before boundary
wrap-around routing.  Odd-$L$ tori require a larger coloring constant.  The
depth remains independent of $L$ at fixed boundary condition and fixed local
compilation pattern.

\paragraph{Number of Trotter steps.}
For second-order Strang--Trotter evolution, the observable error scales as
$\mathcal{O}(\delta t^2)$ at fixed $T$.  The commutator support estimates in
Sec.~\ref{sec:trotter} scale as $L^2$, giving the standard worst-case scaling
\begin{equation}
  N_T \sim T^{3/2}L/\sqrt{\epsilon}
\end{equation}
up to coupling-dependent constants.  The numerical observable-level check in
Sec.~\ref{sec:trotter} gives $N_T=20$ as an adequate $L=4$, $T=10$ benchmark at
the parameters used in the exact diagonalization study.

\paragraph{Total two-qubit gate count and depth.}
With $N_T$ Trotter steps,
\begin{equation}
  G_{\rm 2q}=c_C L^2N_T,
  \qquad
  D_{\rm total}=D_1N_T .
\end{equation}
For $L=4$ and $N_T=20$, this gives $G_{\rm 2q}=2560$, $4480$, or $10240$ and
CNOT-layer depths $640$, $1120$, or $2560$ for the Wilson-core, path-native
full, and routed-star full schedules.

\paragraph{Shot budget.}
A crude circuit-survival model with two-qubit depolarizing rate $p_2$ gives
$\mathcal{F}_{\rm circ}\simeq\exp(-p_2G_{\rm 2q})$.  Combined with the leading
post-selection acceptance estimate from Sec.~\ref{sec:meas},
\begin{equation}
  \eta\simeq \exp[-p_2G_{\rm 2q}-3L^2\bar p]
\end{equation}
for a torus readout-frame error rate $\bar p$.  This expression is a planning
metric rather than a calibrated fidelity model; coherent errors, leakage,
readout correlations, and mitigation change the effective yield.  For the full
path-native $L=4$ schedule, $G_{\rm 2q}=4480$ at $N_T=20$; for the routed-star
full schedule, $G_{\rm 2q}=10240$.  These counts set the required noise level
and motivate the mitigation tests in Sec.~\ref{sec:error}.

\paragraph{Classical-simulation cost.}
For comparison, classical exact diagonalization in the Gauss-law-projected basis
scales as $\mathcal{O}(2^{N_{\rm plaq}})$ in memory and per matrix-free Krylov
step.  The reported exact diagonalization reaches $N_{\rm plaq}=25$ (dimension $2^{25}$) with
matrix-free methods.  Larger square tori require distributed memory,
tensor-network approximations, or quantum hardware.

\section{Exact diagonalization benchmark}
\label{sec:ED}

The principal calculation is the exact simulation of the tube-vs-cap quench on a
$4\times 4$-plaquette spatial \emph{torus} (periodic boundary in both
directions), Hilbert-space dimension $2^{16}$ after Gauss-law projection.
The projection exploits the fact that, in the physical subspace,
$\tau^x_\ell = \prod_{p\ni\ell}\sigma^x_p$, so physical states are
parameterized entirely by the plaquette $\sigma^x_p$ values. The reduced
Hamiltonian is
\begin{align}
H_\mathrm{phys} = \;&\varepsilon_E\, n_-^{(\sigma)}
 + m \sum_\ell \tfrac{1}{2}\bigl(1-\sigma^x_{p_1(\ell)}\sigma^x_{p_2(\ell)}\bigr) \notag\\
 &- \kappa \sum_p \!\!\prod_{p'\text{ nbr of }p}\!\!\!\sigma^x_{p'}
 \;-\; t\sum_p \sigma^z_p,
\label{eq:Hphys}
\end{align}
an Ising-type reduced gauge model on the dual lattice with a 4-body
nearest-neighbour plaquette coupling. For $L_x=L_y=4$ the dimension is
$2^{16}=65536$. The two non-contractible closed strings are placed at the
\emph{asymmetric} separation $y_1=0$, $y_2=1$, where tube and cap differ
strongly in classical energy: $R_\mathrm{tube}$ is a single plaquette row
(4 plaquettes), $R_\mathrm{cap}$ its complement (12 plaquettes). The
symmetric case $y_2-y_1 = L_y/2 = 2$---the classical degeneracy point---is
discussed at the end of the section.

The Hamiltonian was stored as a sparse matrix with $\sim 10^6$ non-zero
entries and time evolved using the Krylov-subspace-based
\texttt{scipy.sparse.linalg.expm\_multiply} routine with step $\delta t=0.2$,
total time $T=30$ (150 steps). Energy conservation is maintained to
$\lesssim 10^{-12}$. 
%The complete run took about 85 seconds on a single CPU.

Initial states: $\ket{\psi_0^\mathrm{tube}} = W[R_\mathrm{tube},
\gamma_1\!\cup\!\gamma_2]\ket{+}^{\otimes N}$ and
$\ket{\psi_0^\mathrm{cap}} = W[R_\mathrm{cap}, \gamma_1\!\cup\!\gamma_2]\ket{+}^{\otimes N}$.
Their initial energies differ by $8\varepsilon_E$. For the straight-boundary geometry used here the initial expectation of the $\kappa$ term vanishes and the off-diagonal $t$ term has zero expectation, giving $E_0^\mathrm{tube}=12.00$ and $E_0^\mathrm{cap}=20.00$ at $\varepsilon_E=m=1$, $t=\kappa=0.4$.

\begin{figure}[!htbp]
\centering
\includegraphics[width=1.0\linewidth]{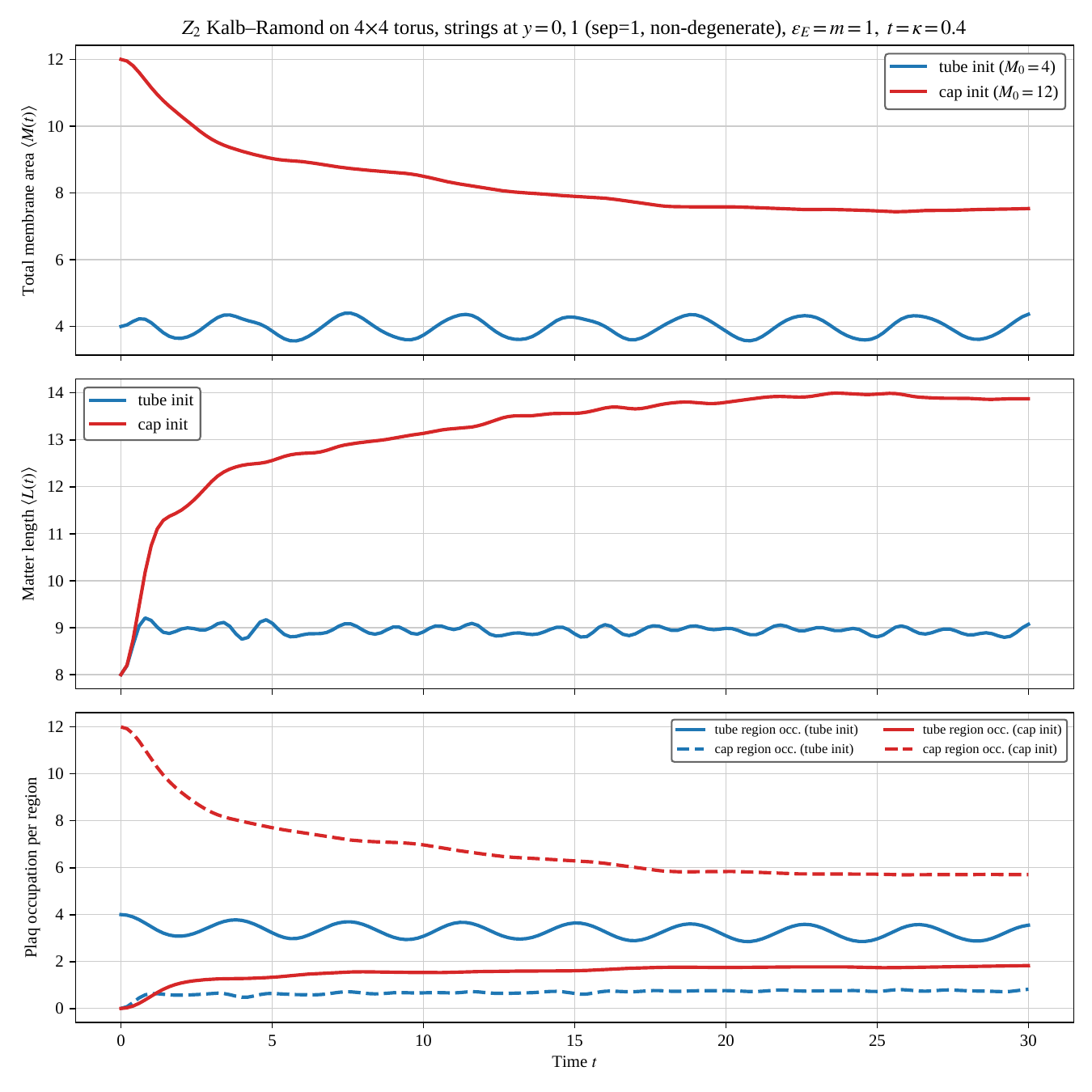}
\caption{Real-time exact diagonalization of the tube-vs-cap quench on a $4\times 4$ spatial
torus (16 plaquette qubits, $2^{16}$ Gauss-law-projected Hilbert space). Two
parallel non-contractible closed string/domain-wall boundaries wrap the periodic
$x$-direction at $y_1=0$ and $y_2=1$ (asymmetric separation $1 < L_y/2$).
Blue: \emph{tube} initial state, electric-flux region on the single plaquette
row between the strings ($M_0=4$). Red: \emph{cap} initial state, electric-flux region on the complement strip of three rows ($M_0=12$). Top: total
occupied-flux area $M(t)$. Middle: total boundary length $L(t)$. Bottom:
per-region plaquette occupation---solid lines track the tube region
(1 row, max 4), dashed lines track the cap region (3 rows, max 12). Blue
curves (tube init): the system is near its classical ground state,
$M(t)\approx 4$ with small oscillations and only mild boundary
nucleation. Red curves (cap init): flux-domain discharge decay is clearly
visible. Total $M$ drops from 12 toward $\approx 7.5$; $L$ grows from 8 to
$\approx 14$ as new closed string/domain-wall loops nucleate in the interior of the
oversized cap flux region; cap-region occupation decays $12\to 6$ while
tube-region occupation grows $0\to 2$, demonstrating that the cap flux region fragments predominantly \emph{in place} through interior closed-loop formation rather than migrating coherently to the tube.}
\label{fig:ED}
\end{figure}

\begin{figure}[!htbp]
\centering
\includegraphics[width=0.5\linewidth]{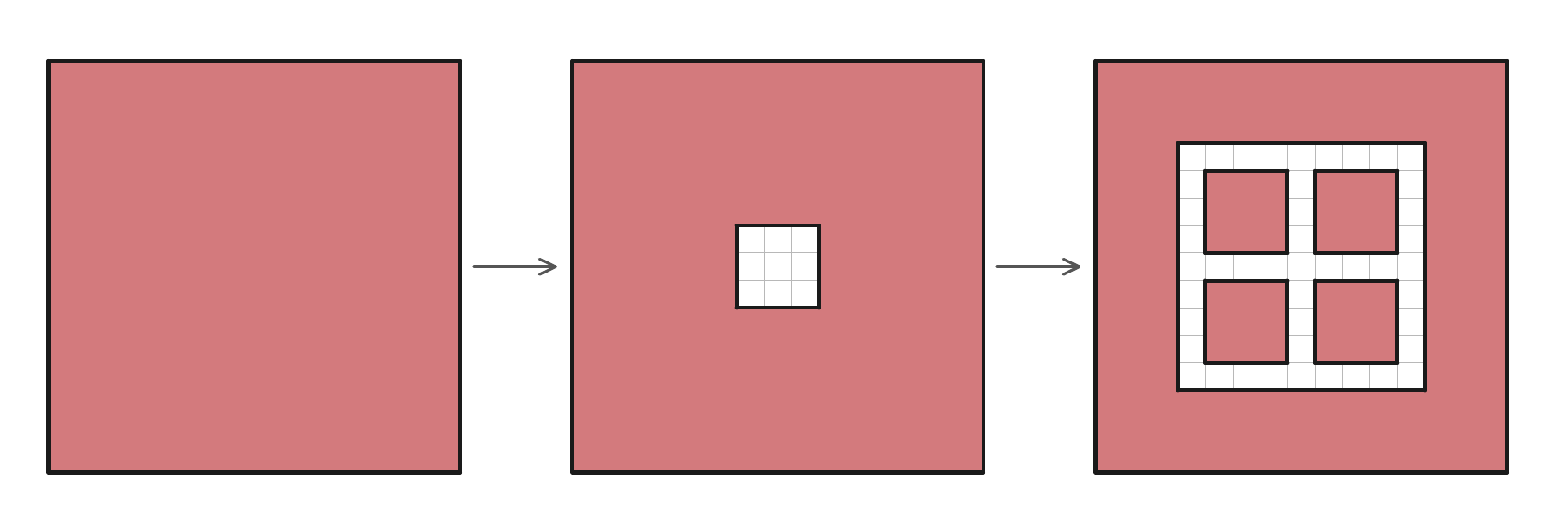}
\caption{Topological cartoon of the cap-discharge cascade resolved by the
exact diagonalization data of Fig.~\ref{fig:ED}. Red plaquettes carry
top-form electric flux ($\sigma^x_p = -1$); white plaquettes are the
dual vacuum. Closed black curves are string/domain-wall boundaries
$\tau^x_\ell = -1$ on plaquette boundaries. Left: initial cap state,
high-flux filling of the cap region. Center: first nucleation
event, a closed string/domain-wall boundary forms in the interior and encloses a
small dual-vacuum region inside the flux. Right: after further
nucleation events the discharged region has grown and contains a
population of smaller closed boundary loops, each enclosing a residual
flux island. The exact diagonalization data show that this fragmentation occurs
\emph{in place}: cap-region occupation decays $12\to 6$ while tube-region
occupation grows only $0\to 2$, consistent with interior Wilson-surface-driven loop formation controlled primarily by $t$, with $\kappa$ modifying the diagonal interaction landscape, rather than coherent translation of the cap.}
\label{fig:cap_cascade_cartoon}
\end{figure}

Figure~\ref{fig:ED} resolves the dynamics of the two initial states
separately. The tube-initial trajectory (blue) is nearly stationary: at
these parameters the tube state is close to the classical ground state of
$H_\mathrm{phys}$, and only small fluctuations around $M\approx 4$ and
$L\approx 9$ develop, reflecting virtual nucleation and annihilation of short closed
loops that quickly re-annihilate. The cap-initial trajectory (red) displays
the flux-domain discharge process: the metastable
12-plaquette cap configuration sheds about 40\% of its electric-flux region
area on a time scale $\sim 5\varepsilon_E^{-1}$, while simultaneously
generating $\sim 6$ additional boundary-string-length units through
dynamical closed-loop nucleation.

Three structural observations follow from the bottom panel. First, most of the
cap-region plaquette deficit is converted locally into additional derived domain-wall boundary length, while a smaller fraction flows into the tube region.
This is \emph{multi-local} flux-domain fragmentation: the large flux domain breaks
through several local boundary-loop formation events rather than by coherent
translation. Second, the tube-region occupation approaches the tube-ground
value without becoming a pure tube state; the late-time state is a fragmented
configuration with the same total energy distributed across many small loops.
Third, the full Hamiltonian energy is conserved to machine precision. The local
boundary-loop dynamics redistributes the initial cap excess among electric-flux region
area, boundary length, and the $H_t$ and $H_\kappa$ interaction energies.

The asymmetric signature $(M_\mathrm{cap}(t)-M_\mathrm{tube}(t))$---the difference between the occupied-flux area starting from cap and from tube initial states---is a sharp probe of the tube-vs-cap energy splitting.  It compares two distinct electric-flux fillings with the same boundary data and is therefore sensitive to the topology of the initial electric-flux region, not only to boundary length.

\paragraph{Physical interpretation: flux-domain fragmentation at fixed
energy.}
The four-panel observation $(M_{\rm tube}, M_{\rm cap}, L_{\rm tube}, L_{\rm cap})$ on
the $4\times 4$ torus has a direct interpretation. The tube-initial trajectory at $\varepsilon_E=m=1$,
$t=\kappa=0.4$ is a near-vacuum configuration: the 4-plaquette tube is
the classical minimum at $E_{\rm tube} = 4\varepsilon_E + 8m = 12$, and
$M_{\rm tube}(t)$ oscillates with amplitude $\lesssim 0.5$ around its
initial value $M=4$ throughout $T=20$. This is the small-amplitude
quantum fluctuation around the classical ground state---virtual closed-loop
boundary-loop formation and annihilation, with negligible real fragmentation. It is
the finite-volume heavy-boundary, or pinned-flux-domain, precursor of the
continuum regime discussed in Ref.~\cite{ReyConfining1991}.

The cap-initial trajectory is qualitatively different: $M_{\rm cap}$
decays monotonically from $12$ to $\approx 7.5$ by $T=20$ (a $37\%$
decrease), while $L_{\rm cap}$ rises from $8$ to $\approx 13.8$ (a
$72\%$ increase). Total energy is conserved to machine precision throughout, but the conserved
quantity is the full Hamiltonian expectation, including the $t$ and $\kappa$
terms:
\begin{equation}
\langle H\rangle(t)
=\langle H_E+H_m+H_\kappa+H_t\rangle(t)
=E_{\rm cap}=20.
\end{equation}
The decay of $M_{\rm cap}$ is therefore not energy loss but redistribution
among electric tension, string length, and interaction energy.  The dominant
visible trend is that the $\varepsilon_E$-tension contribution falls while the
matter-length contribution rises; the remaining difference is carried by the
$H_t$ and $H_\kappa$ expectation values. Microscopically, the oversized
12-plaquette cap flux region breaks apart into multiple smaller disconnected
pieces by Wilson-surface action; each broken interior edge creates a new
closed string/domain-wall loop on its boundary. This is flux-domain fragmentation by local boundary-loop formation: the process acts on a two-dimensional flux region and produces one-dimensional closed boundaries.

\paragraph{Late-time plateau.}
The plateau $M_{\rm cap}(t\to T_{\rm max})\approx7.5$ lies above the tube-ground
value $M=4$.  Energy conservation confines a cap-energy initial state to the
$E=20$ energy shell, so the observed plateau is a finite-volume dephasing
plateau within that shell rather than relaxation to the tube state.  Calling it
a microcanonical value requires an eigenstate-resolved diagonal-ensemble or
longer-time analysis.  In the present $\Z_2$ truncation the plateau represents a
fragmented domain-wall/boundary-string state inside the Gauss-law-projected
plaquette Hilbert space.  A literal string-field condensate requires the
enlarged link Hilbert space discussed in Sec.~\ref{sec:linkboson}.

\paragraph{Finite-size interpretation.}
The larger-volume data test whether the cap-decay signal survives beyond the
smallest torus.  At matched early times, $4\times4$ and $6\times4$ show close
normalized trajectories, while $5\times5$ deviates.  This establishes
persistence of the discharge signal but does not by itself determine a
thermodynamic equation of state.  The next finite-size sequence should hold
aspect ratio and initial energy density fixed, for example comparing
$4\times4$, $6\times6$, and $8\times8$ tori or a matched rectangular sequence.
Such runs separate volume effects from aspect-ratio effects and fix whether the
$4\times4$ plateau is an asymptotic value or a finite-volume transient.

\paragraph{Degeneracy-point check.} The simulation was also run at the
symmetric separation $y_2 - y_1 = L_y/2 = 2$, where classical tube and
cap energies coincide by the exact $L_y/2$-translation symmetry of the
torus. At that point the global observables $M(t)$ and $L(t)$ for tube
and cap initial states are identical to machine precision---a consistency
check confirming both the symmetry of the Hamiltonian and the correctness
of the tube/cap construction. The per-region occupations, which break the
symmetry by labeling, still reveal the coherent tube$\leftrightarrow$cap
mixing at that point.

\subsection{Finite-size scaling: 4$\times$4, 6$\times$4, 5$\times$5}
\label{sec:scaling}

\begin{figure}[!htbp]
\centering
\includegraphics[width=1.0\linewidth]{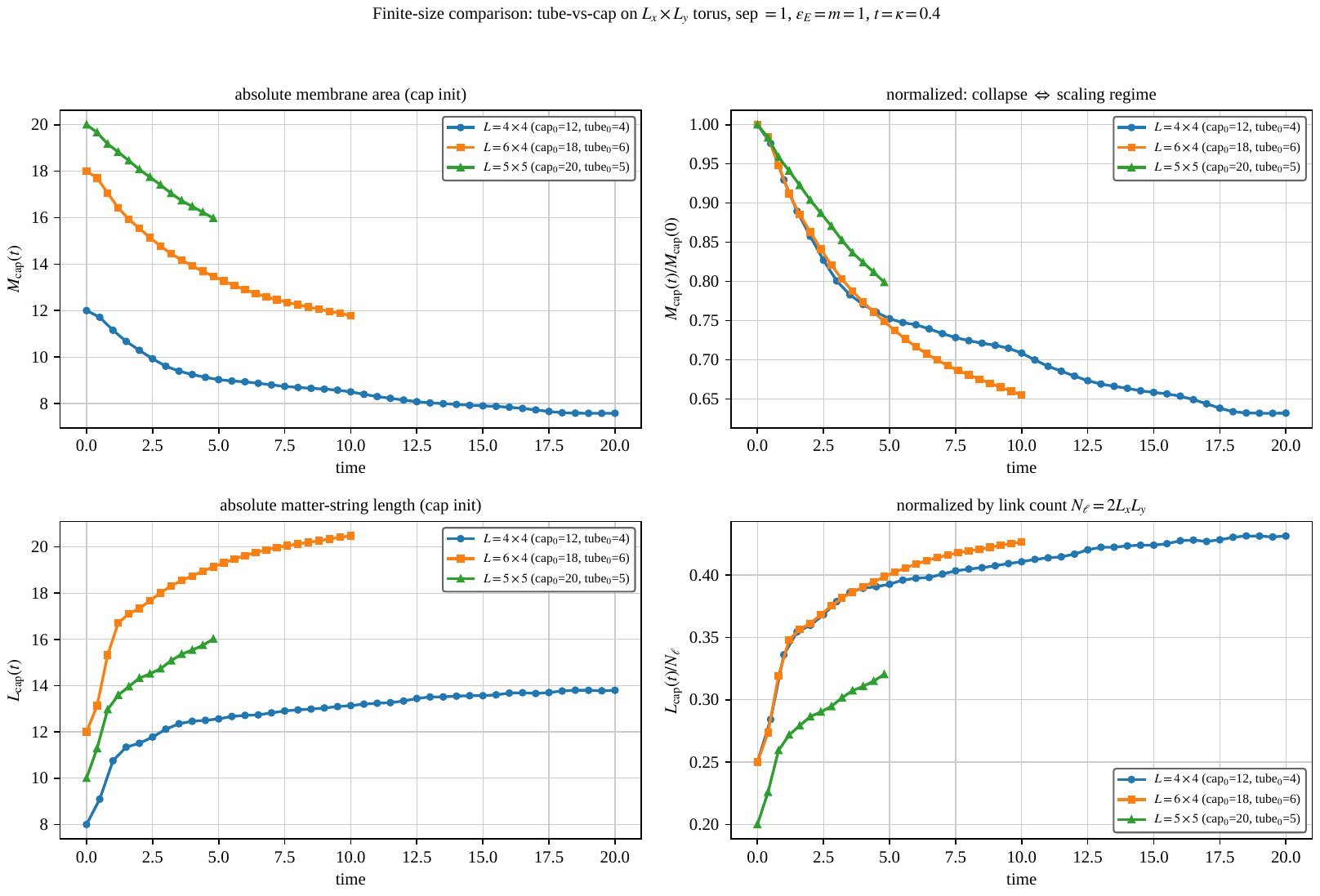}
\caption{Finite-size comparison of the tube-vs-cap quench protocol on
three torus sizes. Top row: total electric occupied-flux area $M_{\rm cap}(t)$
starting from the cap initial state (left, absolute) and normalized by
the initial value (right). Bottom row: total boundary length
$L_{\rm cap}(t)$ (left, absolute) and normalized by the link count
$N_\ell=2L_xL_y$ (right). The 4$\times$4 (blue) and 6$\times$4 (orange)
trajectories, both with $L_x>L_y$, collapse to within $\lesssim 2\%$ on
the normalized scales over the entire accessible time window. The
5$\times$5 (green) trajectory, the only square lattice in the comparison,
sits visibly off the collapsed envelope, decaying $\sim 5$--$7\%$
\emph{slower} on the normalized $M_{\rm cap}$ panel and $\sim 20\%$ lower
on the normalized $L_{\rm cap}$ panel. The deviation is consistent with a mixture of aspect-ratio and volume effects:
the closed-string circumference $L_x$ controls the number of boundary sites
available for local nucleation, while the transverse size controls the cap
interior.  An aspect-ratio-controlled sequence is required to separate these
effects quantitatively.}
\label{fig:scaling}
\end{figure}

The finite-size test repeats the protocol on three lattice sizes: $4\times4$
(16 plaquettes, dim $2^{16}$), $6\times4$ (24 plaquettes, dim $2^{24}$),
and $5\times5$ (25 plaquettes, dim $2^{25}$). All three runs use the
same Hamiltonian parameters $\varepsilon_E=m=1$, $t=\kappa=0.4$, the
same separation $y_2-y_1=1$, and a second-order Strang-Trotter
integrator with $\delta t=0.1$ (4$\times$4) or $\delta t=0.2$ (6$\times$4
and 5$\times$5). The resulting cap-initialization trajectories,
normalized by the initial classical values $M_{\rm cap}(0)$ and the link
count $N_\ell = 2L_xL_y$, are plotted in Fig.~\ref{fig:scaling}.

The qualitative phenomenology is consistent across all three
sizes studied: cap-initialized states monotonically shed $20$--$37\%$ of their
occupied-flux area on a comparable time scale, while tube-initialized states
remain pinned within $\pm 0.5$ of their classical value (Tab.~\ref{tab:scaling}).
The difference $M_{\rm cap}(t)-M_{\rm tube}(t)$ is a robust, sign-definite
signal on every lattice studied. This level of agreement between
4$\times$4 and 6$\times$4 shows that the cap-decay signal persists beyond the
smallest lattice.

\begin{table}[!htbp]
\centering\small
\begin{tabular}{@{}lrccc@{}}
\toprule
$L_x{\times}L_y$ ($t_{\rm end}$) & $N_{p}$ & $M_{\rm cap}(0)$
 & $\frac{M_{\rm cap}(t_{\rm end})}{M_{\rm cap}(0)}$
 & $\frac{L_{\rm cap}(t_{\rm end})}{N_\ell}$ \\
\midrule
$4{\times}4$ ($T{=}20$) & 16 & 12 & 0.63 & 0.43 \\
$4{\times}4$ ($T{=}5$)  & 16 & 12 & 0.75 & 0.39 \\
$6{\times}4$ ($T{=}10$) & 24 & 18 & 0.66 & 0.43 \\
$6{\times}4$ ($T{=}5$)  & 24 & 18 & 0.75 & 0.40 \\
$5{\times}5$ ($T{\approx}5$) & 25 & 20 & 0.80 & 0.32 \\
\bottomrule
\end{tabular}
\caption{Late-time behavior of cap-initialized observables on each
lattice. The $4\times 4$ and $6\times 4$ rectangular runs collapse on
the normalized scale to within a few percent at matched times. The
$5\times 5$ square-lattice run differs systematically by 5--20\% on the
normalized scales.}
\label{tab:scaling}
\end{table}

The $5\times5$ deviation shows that the available data do not yet define a
single-variable finite-size collapse.  Boundary-loop formation is sensitive to the
closed-string circumference $L_x$, while cap fragmentation is also sensitive to
the transverse size and total volume.  A controlled scaling law therefore
requires holding aspect ratio and initial energy density fixed while increasing
$L$.  The classical cost rises rapidly: a $6\times6$ torus has dimension
$2^{36}$ and an $8\times8$ torus has dimension $2^{64}$.  The present data show
that the cap-vs-tube splitting persists as the lattice is enlarged in the
available benchmark geometries.

The computational cost scales as $2^{N_\mathrm{plaq}}$; in the
Gauss-law-projected basis, exact diagonalization is feasible up to $N_\mathrm{plaq} \sim 25$
(dimension $\sim 3\cdot 10^7$) with matrix-free Krylov methods on a
workstation, enabling classical validation of lattices up to $5\times 5$.
Larger targets ($6\times 6$ and beyond) require tensor-network methods,
distributed memory, or quantum hardware.

\subsection{Locating the boundary-proliferation crossover: $m/\varepsilon_E$ sweep}
\label{sec:m-eps-sweep}

The benchmark of Sec.~\ref{sec:ED} sets $m/\varepsilon_E = 1$, i.e.\
classical boundary line tension equal to plaquette-flux energy. The two limits
$m/\varepsilon_E \to 0$ and $m/\varepsilon_E \to \infty$ realize,
respectively, a light-boundary proliferation regime and a heavy-boundary confined regime.  These are the finite-$\Z_2$, finite-volume precursors of the continuum string-Higgs and confining regimes of the Kalb--Ramond construction~\cite{ReyHiggs1989,ReyConfining1991}. To locate the crossover between them on the $4\times 4$ torus, a sweep was performed
$m/\varepsilon_E \in \{0.25, 0.5, 1, 2, 4\}$ at fixed
$t = \kappa = 0.4$, evolving the cap initial state to $T = 20$.

\begin{table}[!htbp]
\centering
\begin{tabular}{@{}rrrrr@{}}
\toprule
$m/\varepsilon_E$ & $E_{\rm cap}(0)$ & $M_{\rm cap}(T)$ & $\Delta M / M_0$
& $-\partial_t M|_0$ \\
\midrule
0.25 & 14.0 & 8.94  & 25.5\% & 1.78  \\
0.5  & 16.0 & 8.17  & 32.0\% & 1.73  \\
1.0  & 20.0 & 7.59  & 36.8\% & 0.94  \\
1.25 & 22.0 & 8.11  & 32.4\% & 0.58  \\
1.5  & 24.0 & 9.02  & 24.8\% & 0.31  \\
1.75 & 26.0 & 10.63 & 11.4\% & 0.10  \\
2.0  & 28.0 & 10.16 & 15.4\% & $-0.07$ \\
4.0  & 44.0 & 12.21 & $-1.7\%$ & $-0.45$ \\
\bottomrule
\end{tabular}
\caption{Cap-decay observables as a function of the boundary-line-tension to
plaquette-flux-energy ratio on the $4\times 4$ torus, at $t=\kappa=0.4$,
$T_{\rm max}=20$. $E_{\rm cap}(0)$ is the initial energy
$E = 12\,\varepsilon_E + 8\,m$; $M_{\rm cap}(T)$ is the
late-time occupied-flux area; $\Delta M/M_0 = (M_0 - M(T))/M_0$ is the
fractional decay; $-\partial_t M|_0$ is the initial decay rate from
a linear fit to $t \in [0, 2]$. The five bisection ratios
$\{1.25, 1.5, 1.75\}$ refine the location of the dynamical crossover
and bracket the zero crossing between $1.75$ ($-\partial_t M|_0 = +0.10$) and
$2.0$ ($-\partial_t M|_0 = -0.07$). Energy is conserved to
$|\Delta E| < 5\cdot 10^{-12}$ on every run.}
\label{tab:m-eps-sweep}
\end{table}

\begin{figure}[!htbp]
\centering
\includegraphics[width=1.0\linewidth]{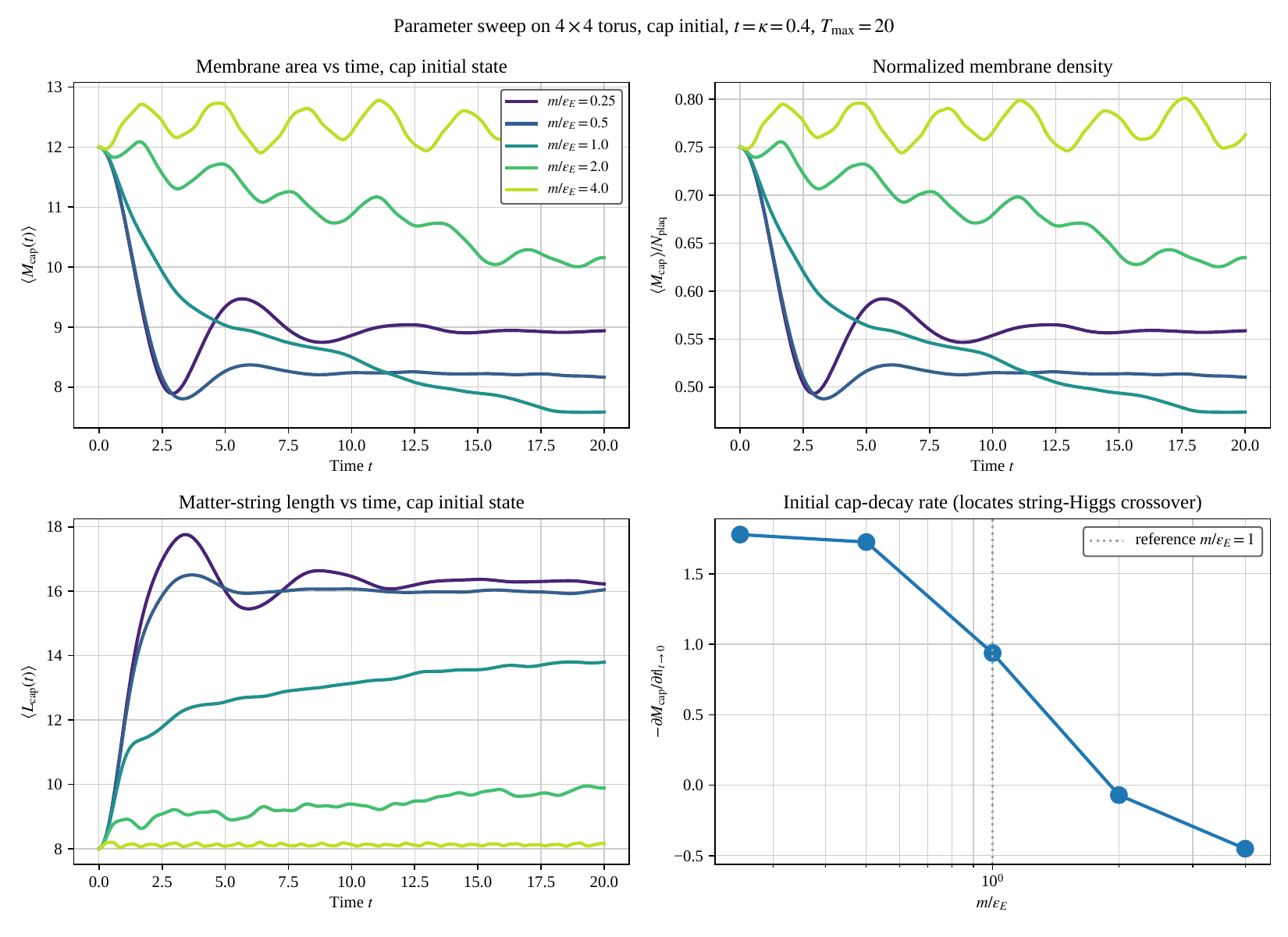}
\caption{$m/\varepsilon_E$ sweep on the $4\times 4$ torus. Top-left:
$\langle M_{\rm cap}(t)\rangle$ for ratios $\{0.25,0.5,1,2,4\}$.
Top-right: same quantity normalized by $N_{\rm plaq}=16$. Bottom-left:
boundary length $\langle L_{\rm cap}(t)\rangle$.
Bottom-right: initial cap-decay rate $-\partial_t M_{\rm cap}|_{t\to 0}$
versus $m/\varepsilon_E$, locating the dynamical boundary-proliferation crossover
between $m/\varepsilon_E \approx 1$ and $m/\varepsilon_E \approx 2$ at
these $(t, \kappa)$ values.}
\label{fig:m-eps-sweep}
\end{figure}

The results are summarized in Table~\ref{tab:m-eps-sweep} and
Fig.~\ref{fig:m-eps-sweep}. They exhibit a finite-volume dynamical crossover
between two qualitatively distinct regimes:

\emph{Light-boundary proliferation regime,
$m/\varepsilon_E \le 1$.} The cap flux region fragments rapidly: $M_{\rm cap}$
drops by 25--37\% over $T = 20$, while the boundary length
$L_{\rm cap}$ approximately doubles. The initial decay rate
$-\partial_t M_{\rm cap}|_{0} \approx 1.0$--$1.8$ scales inversely with
$m$, consistent with the Wilson-surface boundary-nucleation amplitude $t$
unsuppressed by the boundary line-energy cost. This is the $\Z_2$ dynamical signature of boundary proliferation: light closed boundaries form readily in the flux-region interior, breaking the oversized cap into many small flux-domain--boundary composites.

\emph{Heavy-boundary confined regime, $m/\varepsilon_E \ge 2$.}
Fragmentation is suppressed: at $m/\varepsilon_E = 2$, $M_{\rm cap}$
decays by only 15\%; at $m/\varepsilon_E = 4$, the cap is essentially
stable ($M$ varies within $\pm 0.3$ of its initial value 12, a quantum
fluctuation amplitude). The fitted initial decay rate changes sign between
$m/\varepsilon_E=1.75$ and $2.0$ and becomes mildly negative
(quantum fluctuation now \emph{increases} $M$ slightly) at
$m/\varepsilon_E = 4$. Closed boundaries are then too costly to nucleate efficiently, so the flux region remains pinned near its initial topology.

\emph{Crossover location.} A coarse sweep over
$m/\varepsilon_E \in \{0.25, 0.5, 1, 2, 4\}$ brackets the crossover between
$1$ and $2$. A finer bisection adding the intermediate ratios
$\{1.25, 1.5, 1.75\}$ resolves the cap-decay-rate zero-crossing more
tightly: the initial decay rates $-\partial_t M_{\rm cap}|_0$ go through
$0.94, 0.58, 0.31, 0.10, -0.07$ at $m/\varepsilon_E = 1.0, 1.25, 1.5,
1.75, 2.0$ respectively, locating the dynamical crossover on the
$4\times 4$ lattice at
\begin{equation}
\label{eq:crossover}
  (m/\varepsilon_E)_c \;\approx\; 1.89
\end{equation}
by linear interpolation in $\log(m/\varepsilon_E)$ between the bracket
$1.75$ (rate $+0.10$) and $2.00$ (rate $-0.07$). The benchmark
parameters $(\varepsilon_E, m) = (1, 1)$ used elsewhere in this paper sit
on the boundary-proliferating side of this finite-volume crossover, which is
the appropriate regime for observing dynamical fragmentation. The late-time occupied-flux area
$\langle M_{\rm cap}(T=20)\rangle$ shows the same crossover non-monotonically:
it drops from 12 to a minimum of $\sim 7.6$ at $m/\varepsilon_E = 1.0$,
then rises through the crossover to recover near 12 at
$m/\varepsilon_E = 4$. The non-monotonicity at $m/\varepsilon_E \approx
1.75$--$2.0$ reflects partial reversibility of the fragmentation:
heavier boundaries are created with reduced finite-volume excursions. Mapping out the full
$(m/\varepsilon_E, t/\varepsilon_E, \kappa/\varepsilon_E)$ dynamical diagram is
a direct next measurement.  A parametric sweep re-tunes the Trotter
coefficients and reuses the same circuit topology.

\section{Error mitigation}
\label{sec:error}

The error-mitigation strategy separates errors that leave the Gaus-law sector from
errors that remain inside it.  This separation is the central coding distinction in Gauss-law QED: gauge-violating faults carry a local constraint syndrome, whereas gauge-preserving faults act within the physical code space and require additional protection~\cite{Wiebe2023,Wiebe2026}. The first class is detected directly by the
link constraints at readout.  The second class changes physical observables and
must be handled by ordinary calibration and extrapolation methods.

\emph{(i) Gauge-sector post-selection.}  The final electric-basis bitstring
contains all plaquette and link eigenvalues needed to evaluate every $\mathcal G_\ell$.
Shots violating any measured constraint are discarded.  This step is exact as a
classical filter on readout bitstrings; it does not require additional quantum
measurements beyond the global $X$-basis readout.

\emph{(ii) Pauli twirling.}  Random Pauli dressing of two-qubit gates converts
coherent over-rotations into an effective stochastic Pauli channel.  This is
especially relevant for the parity-accumulation circuits in $H_t$ and routed
$H_\kappa$.

\emph{(iii) Zero-noise extrapolation.}  Runs at several folded two-qubit-gate
noise levels, for example $\lambda\in\{1,2,3\}$, estimate the zero-noise limit
of $M(t)$ and $L(t)$.  The extrapolation model must be fitted and validated on
small lattices where noiseless simulation is available.

\emph{(iv) Readout-error mitigation.}  A calibrated confusion matrix, either
factorized or correlated over local neighborhoods, is applied before or jointly
with Gauss-law-sector filtering.  Because the Gaus-law constraints are parity checks,
correlated readout errors near a plaquette boundary are the relevant failure
mode.

\emph{(v) Observable renormalization.}  Single-qubit $X$ observables can be
renormalized using identity and echo circuits of the same depth.  This corrects
smooth decay envelopes but does not replace the Gauss-law-sector filter.

\emph{(vi) Gauge-aware echo layers.}  Products of commuting Gauss-law generators may
be inserted as digital echo operations when the hardware schedule contains idle
windows.  These layers act trivially on the physical subspace.  They are useful
only when their additional gates reduce more error than they introduce, so they
belong to the device-level optimization rather than the baseline circuit count.

The mitigation target is therefore operational: reproduce the exact-diagonalization-calibrated
observables $M(t)$ and $L(t)$ on small lattices after applying the same
post-selection and extrapolation pipeline.  No fixed percentage-bias claim is
assumed without a device-specific noise model.

\subsection{Numerical feasibility estimate on a $2\times 2$ torus}
\label{sec:noisy}

The noisy-circuit check simulates the full Strang--Trotter circuit with
depolarizing gate noise on a $2\times2$ torus: 4 plaquette qubits, 8 link
qubits, and a Gauss-law-projected physical subspace of dimension $2^4=16$.  The
noise model uses $p_1=10^{-4}$ and
$p_2\in\{5\times10^{-4},2\times10^{-3},5\times10^{-3}\}$.  The reported run
uses $\delta t=0.5$, $N_T=6$, and $N_{\rm shot}=150$ noisy Monte Carlo
trajectories against a noiseless Strang--Trotter reference.

The measured final-time observable bias
$|\Delta M_{\rm cap}(T)|/M_{{\rm cap},{\rm ref}}(T)$ is
\begin{center}
\begin{tabular}{c|ccc}
 $p_2$ & raw & Gauss post-sel. & post-sel. survival \\ \hline
 $5\times 10^{-4}$ & $4.2\%$ & $24\%$ & $77\%$ \\
 $2\times 10^{-3}$ & $13\%$ & $57\%$ & $45\%$ \\
 $5\times 10^{-3}$ & $23\%$ & $87\%$ & $13\%$ \\
\end{tabular}
\end{center}

This table has a clear interpretation.  At $L=2$, Gauss-law post-selection
amplifies the bias because the cap state lies at the edge of the physical
spectrum: depolarizing noise drives the accepted physical component toward the
uniform physical-sector mean $\overline M=2$, whereas the cap initial state has
$M_0=4$.  The result is not evidence against post-selection; it shows that
post-selection removes leakage but does not correct physical-sector mixing.

The larger-lattice behavior must be calibrated directly.  The physical fraction
of the full qubit Hilbert space scales as $2^{-2L^2}$ on the torus, so generic
uncorrelated errors increasingly produce leakage as $L$ grows.  This scaling
supports the expectation that Gauss-law filtering becomes more useful beyond the
$2\times2$ test, but the crossover point is a measured quantity.  An $L=3$
calibration run is the appropriate next noisy-circuit benchmark before any
production $L=4$ run.

\section{Discussion and outlook}
\label{sec:discussion}

The exact diagonalization results establish the real-time dynamics of the complete
Gauss-law-projected $\Z_2$ model. The independent degrees of freedom are plaquette
electric fluxes. Closed link loops are reconstructed exactly from the evolving
plaquette state and therefore provide dynamical boundary observables without
introducing a separate string-matter Hilbert space. This is the correct
interpretation of the tube--cap quench, the boundary-length growth, and the
finite-volume crossover.

The unreduced plaquette-plus-link representation remains valuable for digital
implementation. It realizes every reduced plaquette flip as the local
Gauss-law-preserving operator
$\sigma_p^z\prod_{\ell\in\partial p}\tau_\ell^z$ and exposes the link Gauss
constraints for verification. The reduced and unreduced formulations describe
the same ideal physical trajectories: the former minimizes the Hilbert space
for exact diagonalization, while the latter trades redundant qubits for local checks and local
circuit structure.
This trade between a minimal reduced register and a redundant local gauge-covariant encoding is the same resource trade exploited by Gauss-law quantum codes~\cite{Wiebe2023,Wiebe2026}. The present protocol uses the redundant encoding for local implementation and syndrome verification, but does not yet encode the complete evolution fault tolerance. 

The top-form interpretation is correspondingly precise. The cap state is a
high-occupation electric-flux configuration, and its relaxation redistributes
plaquette flux and derived boundary length at fixed total energy. This is a
lower-dimensional Hamiltonian model of top-form flux discharge. It does not
include gravitational back-reaction, and it does not yet realize an independent
continuum string field.

Several extensions follow directly.

\emph{Higher-form check complexes and qLDPC extensions.}
Section~\ref{sec:qec-structure} separates the two coding roles of the general
Hamiltonian.  The present matter-coupled $p=d=2$ model provides a redundant
gauge encoding with local Gauss-law syndromes, while a pure theory in
$d\ge p+1$ supplies both the $X$-type Gauss-law checks and the $Z$-type magnetic
checks of a homological CSS complex.  On a bounded-degree cellulation this is
an LDPC check algebra; its rate and distance are controlled by the homology and
systoles of the chosen complex.  Replacing a Euclidean lattice by a hyperbolic,
expander, or product complex therefore defines a separate route toward
higher-form qLDPC Hamiltonians.  The central open problem is to preserve local
Wilson dynamics while obtaining growing distance.  That problem is independent
of the exact diagonalization specialization and motivates a dedicated QEC study.

\emph{Dynamical string matter beyond $\Z_2$: link-boson formulation.}
\label{sec:linkboson}
The next extension is to promote the hard-core link variable to an
integer-valued oriented string occupation.  Replace each link qubit by a
truncated bosonic or qudit mode with number $n_\ell\in\mathbb{Z}$ (or a finite
range approximation to it), counting the net oriented number of string strands
on $\ell$.  Closedness of the string configuration is imposed by the
vertex-divergence constraint
\begin{equation}
\label{eq:vertex-div-free}
  \partial_v n\equiv \sum_{\ell:v\in\partial\ell}
  \varepsilon_{\ell v} n_\ell =0\qquad\forall v .
\end{equation}
The one-form Gauss-law then relates the string charge on a link to the oriented
difference of the two adjacent plaquette electric fields,
\begin{equation}
\label{eq:boson-gauss}
  \sum_{p\supset\ell}s_{\ell p}E_p=q\,n_\ell .
\end{equation}
This is the $U(1)$ version of the $\Z_2$ constraint
$\tau^x_\ell=\prod_{p\ni\ell}\sigma^x_p$, but now the link occupation has a
larger independent spectrum.

Let $U_p=e^{iB_p}$ be the plaquette gauge transporter and let
$a_\ell^\dagger,a_\ell$ raise and lower the oriented link occupation.  Define
\begin{equation}
  R_{\ell p}=\begin{cases}
    a_\ell^\dagger, & s_{\ell p}=+1,\\
    a_\ell, & s_{\ell p}=-1,
  \end{cases}
\end{equation}
with the convention reversed if the orientation of the plaquette is reversed.
A schematic local gauge-invariant boundary creation operator is then
\begin{equation}
  \mathcal{O}^{\rm bos}_p
  = U_p\prod_{\ell\in\partial p}R_{\ell p}
  + U_p^\dagger\prod_{\ell\in\partial p}R_{\ell p}^\dagger .
\end{equation}
The corresponding bosonic Wilson-surface Hamiltonian is
\begin{equation}
\label{eq:Ht-boson}
  H_t^{\rm bos}=-t\sum_p \mathcal{O}^{\rm bos}_p .
\end{equation}
This is the genuine number-changing version of the $\Z_2$ 5-body operator.  In
such a model the loop functional $\Phi[\gamma]$ can be represented by amplitudes
of closed occupation patterns satisfying Eq.~\eqref{eq:vertex-div-free}, and
joining/splitting processes arise from repeated local boundary moves.  Mapping
this enlarged model onto the continuum string-Higgs construction of
Ref.~\cite{ReyHiggs1989} is a separate continuum-limit problem beyond the $\Z_2$ benchmark.

\emph{Hardware feasibility of the link-boson construction.}  The operator in
Eq.~\eqref{eq:Ht-boson} is a non-Gaussian multi-mode interaction.  A practical
near-term route is a $\Z_N$ qudit truncation on links with $N=4$ or $8$ before
moving to oscillator modes.  For an $L=3$ open plaquette patch the geometry has
9 plaquette gauge variables and 24 link variables; a qudit or truncated-boson
implementation is therefore already a mixed qubit--qudit problem of moderate
width.  The Meth--Ringbauer qudit architecture~\cite{Meth2025Qudit} and the
Joshi et al. qudit gauge-theory construction~\cite{Joshi2025} are the natural
hardware references for this stage.

\emph{Non-abelian generalization.} Non-abelian two-form gauge theories and
2-group gauge theories admit lattice representations with boundary data on
links, along the Chan--Paton line of Ref.~\cite{ReySugino2010}.  Extending the
present abelian $\Z_2$ construction in that direction requires replacing the
Pauli strings by non-commuting boundary representation data.

\emph{Static string potential and confinement diagnostics.} Beyond the quench
protocol, the same qubit layout supports adiabatic preparation of states
carrying fixed Wilson-surface probes at two sizes.  The extracted flux-area coefficient
$\sigma_\mathrm{flux}(R,\gamma)$ as a function of region shape gives the
area-scaling diagnostic for the heavy-boundary regime and its breakdown in the
boundary-proliferating regime.

\emph{Phase-diagram sweep across $m/\varepsilon_E$.}  Section~\ref{sec:m-eps-sweep}
gives a finite-volume sweep at fixed $t=\kappa=0.4$.  The next sweep
will vary $t/\varepsilon_E$ and $\kappa/\varepsilon_E$ independently.  The
role of $t$ is to generate local boundary-changing transitions; the role of $\kappa$ is a
diagonal link-electric plaquette interaction in the electric basis.  Separating
these effects is necessary before assigning continuum phase-language to the
finite $\Z_2$ crossover.

\emph{Continuum limit.}  Increasing $k$ in a $\Z_k$ truncation approaches the
compact $U(1)$ rotor theory and restores orientation data absent in $\Z_2$.
For $k>2$ the Wilson-surface operator must include the boundary orientation
factors explicitly; the unsigned product is special to $\Z_2$ because
$V_\ell^{-1}=V_\ell$.  A scaling sequence $k=2,3,4,5$ on a fixed lattice tests
which cap-decay observables survive the approach to the compact $U(1)$ theory.

\emph{Trotter-step compression.} Three routes can reduce the per-step CNOT depths quoted in
Sec.~\ref{sec:complexity}: variational
Trotter optimization, fourth-order Suzuki--Trotter splittings, and hybrid
Trotter--Richardson extrapolation.  Each trades circuit depth against gate
count or shot budget, and the optimum depends on the device error profile at
run time.

\emph{Coupling to $(2+1)$-dimensional quantum gravity.}  The Brown--Teitelboim correspondence of
Sec.~\ref{sec:BT} concerns the microscopic flux-discharge Hamiltonian only.  A
full cosmological-constant neutralization model requires gravitational
back-reaction.  Coupling the lattice Hamiltonian to a discretized $(2+1)$-dimensional
quantum gravity sector is a longer-term extension beyond the present hardware target.

The immediate experimental target is the $L=4$ quench: 48 data qubits for the
torus topology used in exact diagonalization, or 56 data qubits for the open-patch variant.  The
technical core is the Gauss-law-preserving local Wilson-surface operator, the
routing-aware treatment of $H_\kappa$, and the per-link post-selection rule.
Demonstrating the cap-vs-tube asymmetry on hardware would realize a real-time
microscopic simulation of top-form electric-flux discharge in a higher-form
lattice gauge theory; the enlarged $\Z_k$ or link-boson versions then test the
same mechanism with oriented and multi-occupancy string sectors.

\acknowledgments
I acknowledge discussions with Joshua Lin, Enrique Rico Ortega, Tommaso Rainaldi, Felix Ringer, Xiaojun Yao, and other participants of the workshops at the CERN Theory Division (August 2025, Switzerland), the Institute for Nuclear Theory (December 2025, USA), and the Mainz Institute for Theoretical Physics (April 2026, Germany). Part of this work was performed while I was visiting the Institute for Pure and Applied Mathematics (IPAM, USA), the Simons Institute for the Theory of Computing (SIfTC, USA), and the Fields Institute for Research in Mathematical Sciences (FIRMS, Canada). This work was supported in part by the DFG Cluster of Excellence PRISMA+ (Project ID 39083469), by the U.S. Department of Energy through INT, by the U.S. National Science Foundation through IPAM and SIfTC, by the Simons Foundation through SIfTC, and by the National Research Foundation of Korea (NRF) (RS-2021-NR060112) and Kwangwoon University.


\begin{thebibliography}{99}

\bibitem{KalbRamond}
M.\ Kalb and P.\ Ramond, \textit{Classical direct interstring action},
Phys.\ Rev.\ D \textbf{9}, 2273 (1974).

\bibitem{ReyHiggs1989}
S.-J.\ Rey,
\textit{Higgs mechanism for Kalb--Ramond gauge field},
Phys.\ Rev.\ D \textbf{40}, 3396 (1989).

\bibitem{ReyConfining1991}
S.-J.\ Rey,
\textit{Confining phase of superstrings and axionic strings},
Phys.\ Rev.\ D \textbf{43}, 526 (1991).

\bibitem{ReySugino2010}
S.-J.\ Rey and F.\ Sugino,
\textit{A nonperturbative proposal for nonabelian tensor gauge theory and
dynamical quantum Yang--Baxter maps},
\href{https://arxiv.org/abs/1002.4636}{arXiv:1002.4636} [hep-th] (2010).

\bibitem{Wiebe2023} A.~Rajput, A.~Roggero, and N.~Wiebe, \textit{Quantum error correction with gauge symmetries,} npj Quant. Info. \textbf{9}, 41 (2023). 

\bibitem{Wiebe2026} L.~Spagnoli, A.~Roggero, and N.~Wiebe, \textit{Fault-tolerant simulation of Lattice Gauge Theories with gauge covariant codes}, Quantum {\bf 10}, 1968  (2026).

\bibitem{BrownTeitelboim1987}
J.\ D.\ Brown and C.\ Teitelboim,
\textit{Dynamical neutralization of the cosmological constant},
Phys.\ Lett.\ B \textbf{195}, 177 (1987).

\bibitem{BrownTeitelboim1988}
J.\ D.\ Brown and C.\ Teitelboim,
\textit{Neutralization of the cosmological constant by membrane creation},
Nucl.\ Phys.\ B \textbf{297}, 787 (1988).

\bibitem{BanulsEPJD2020}
M.\ C.\ Ba\~nuls \textit{et al.},
\textit{Simulating lattice gauge theories within quantum technologies},
Eur.\ Phys.\ J.\ D \textbf{74}, 165 (2020),
\href{https://arxiv.org/abs/1911.00003}{arXiv:1911.00003}.

\bibitem{DiMeglio2024}
A.\ Di Meglio \textit{et al.},
\textit{Quantum computing for high-energy physics: State of the art and
challenges},
PRX Quantum \textbf{5}, 037001 (2024).

\bibitem{HebenstreitBergesGelfand2013}
F.\ Hebenstreit, J.\ Berges, and D.\ Gelfand,
\textit{Real-time dynamics of string breaking},
Phys.\ Rev.\ Lett.\ \textbf{111}, 201601 (2013),
\href{https://arxiv.org/abs/1307.4619}{arXiv:1307.4619}.

\bibitem{Pichler2016}
T.\ Pichler, M.\ Dalmonte, E.\ Rico, P.\ Zoller, and S.\ Montangero,
\textit{Real-time dynamics in $U(1)$ lattice gauge theories with tensor
networks},
Phys.\ Rev.\ X \textbf{6}, 011023 (2016),
\href{https://arxiv.org/abs/1505.04440}{arXiv:1505.04440}.

\bibitem{Rigobello2021}
M.\ Rigobello, S.\ Notarnicola, G.\ Magnifico, and S.\ Montangero,
\textit{Entanglement generation in (1+1)D QED scattering processes},
Phys.\ Rev.\ D \textbf{104}, 114501 (2021),
\href{https://arxiv.org/abs/2105.03445}{arXiv:2105.03445}.

\bibitem{PapaefstathiouBanuls2025}
I.\ Papaefstathiou, J.\ Knolle, and M.\ C.\ Ba\~nuls,
\textit{Real-time scattering in the lattice Schwinger model},
Phys.\ Rev.\ D \textbf{111}, 014504 (2025),
\href{https://arxiv.org/abs/2402.18429}{arXiv:2402.18429}.

\bibitem{BelyanskyDavoudiGorshkov2024}
R.\ Belyansky, S.\ Whitsitt, N.\ Mueller, A.\ Fahimniya, E.\ R.\ Bennewitz,
Z.\ Davoudi, and A.\ V.\ Gorshkov,
\textit{High-energy collision of quarks and mesons in the Schwinger model:
From tensor networks to circuit QED},
Phys.\ Rev.\ Lett.\ \textbf{132}, 091903 (2024),
\href{https://arxiv.org/abs/2307.02522}{arXiv:2307.02522}.

\bibitem{Lin2024}
J.~Lin, D.~Luo, X.~Yao, and P.E.~Shanahan, \textit{Real-time dynamics of the Schwinger model as an open quantum system with Neural Density Operators}, J.\ High\ Energy \ Phys. \textbf{ 2024}, 1 (2024).

\bibitem{Ale2026}
V.~Ale, T.~Rainaldi, E.~Rico, F.~Ringer, and G.~Siopsis. \textit{Simulating quantum electrodynamics in 2+ 1 dimensions with qubits and qumodes}, J.\ High\ Energy \ Phys. \textbf{2026}, 122 (2026).

\bibitem{Nguyen2022Schwinger}
N.\ H.\ Nguyen, M.\ C.\ Tran, Y.\ Zhu, A.\ M.\ Green, C.\ H.\ Alderete,
Z.\ Davoudi, and N.\ M.\ Linke,
\textit{Digital quantum simulation of the Schwinger model and symmetry
protection with trapped ions},
PRX Quantum \textbf{3}, 020324 (2022),
\href{https://arxiv.org/abs/2112.14262}{arXiv:2112.14262}.

\bibitem{DavoudiHsiehKadam2024}
Z.\ Davoudi, C.-C.\ Hsieh, and S.\ V.\ Kadam,
\textit{Scattering wave packets of hadrons in gauge theories: Preparation
on a quantum computer},
Quantum \textbf{8}, 1520 (2024),
\href{https://arxiv.org/abs/2402.00840}{arXiv:2402.00840}.

\bibitem{DavoudiHsiehKadam2025}
Z.\ Davoudi, C.-C.\ Hsieh, and S.\ V.\ Kadam,
\textit{Quantum computation of hadron scattering in a lattice gauge
theory},
\href{https://arxiv.org/abs/2505.20408}{arXiv:2505.20408} (2025).

\bibitem{Bennewitz2025}
E.\ R.\ Bennewitz, B.\ Ware, A.\ Schuckert, A.\ Lerose, F.\ M.\ Surace,
R.\ Belyansky, W.\ Morong, D.\ Luo, A.\ De, K.\ S.\ Collins, O.\ Katz,
C.\ Monroe, Z.\ Davoudi, and A.\ V.\ Gorshkov,
\textit{Simulating meson scattering on spin quantum simulators},
Quantum \textbf{9}, 1773 (2025),
\href{https://arxiv.org/abs/2403.07061}{arXiv:2403.07061}.

\bibitem{Schuhmacher2025}
J.\ Schuhmacher, G.-X.\ Su, J.\ J.\ Osborne, A.\ Gandon, J.\ C.\ Halimeh,
and I.\ Tavernelli,
\textit{Observation of hadron scattering in a lattice gauge theory on a
quantum computer},
\href{https://arxiv.org/abs/2505.20387}{arXiv:2505.20387} [quant-ph]
(2025).

\bibitem{CobosZ2Higgs}
J.\ Cobos, J.\ Fraxanet, C.\ Benito, F.\ di Marcantonio, P.\ Rivero,
K.\ Kap\'as, M.\ A.\ Werner, \"O.\ Legeza, A.\ Bermudez, and E.\ Rico,
\textit{Real-time dynamics in a $(2+1)$-D gauge theory: The stringy
nature on a superconducting quantum simulator},
\href{https://arxiv.org/abs/2507.08088}{arXiv:2507.08088} [quant-ph]
(2025).

\bibitem{Cochran2025Nature}
T.\ A.\ Cochran, B.\ Jobst, E.\ Rosenberg, Y.\ D.\ Lensky, G.\ Gyawali,
N.\ Eassa, M.\ Will, D.\ Abanin, R.\ Acharya, L.\ A.\ Beni \textit{et al.},
\textit{Visualizing dynamics of charges and strings in (2+1)D lattice gauge
theories},
Nature \textbf{642}, 315 (2025),
\href{https://arxiv.org/abs/2409.17142}{arXiv:2409.17142}.

\bibitem{GonzalezCuadra2025Nature}
D.\ Gonz\'alez-Cuadra \textit{et al.},
\textit{Observation of string breaking on a $(2+1)$D Rydberg quantum
simulator},
Nature \textbf{642}, 321 (2025).

\bibitem{BoussoPolchinski2000}
R.\ Bousso and J.\ Polchinski,
\textit{Quantization of four-form fluxes and dynamical
neutralization of the cosmological constant},
JHEP \textbf{06}, 006 (2000),
\href{https://arxiv.org/abs/hep-th/0004134}{hep-th/0004134}.

\bibitem{HenneauxTeitelboim1984}
M.\ Henneaux and C.\ Teitelboim,
\textit{The cosmological constant as a canonical variable},
Phys.\ Lett.\ B \textbf{143}, 415 (1984).

\bibitem{HuangShadows}
H.-Y.\ Huang, R.\ Kueng, and J.\ Preskill,
\textit{Predicting many properties of a quantum system from very few
measurements},
Nat.\ Phys.\ \textbf{16}, 1050 (2020).

\bibitem{NielsenChuang}
M.\ A.\ Nielsen and I.\ L.\ Chuang,
\textit{Quantum computation and quantum information},
Cambridge Univ.\ Press (2010).

\bibitem{Meth2025Qudit}
M.\ Meth, J.\ Zhang, J.\ F.\ Haase, C.\ Edmunds, L.\ Postler,
A.\ Steiner, A.\ Jena, L.\ Dellantonio, R.\ Blatt, P.\ Zoller,
T.\ Monz, P.\ Schindler, C.\ Muschik, and M.\ Ringbauer,
\textit{Simulating two-dimensional lattice gauge theories on a qudit
quantum computer},
Nat.\ Phys.\ \textbf{21}, 570 (2025).

\bibitem{Joshi2025}
R.\ Joshi, J.\ C.\ Louw, M.\ Meth, J.\ J.\ Osborne, K.\ Mato,
G.-X.\ Su, M.\ Ringbauer, and J.\ C.\ Halimeh,
\textit{Probing hadron scattering in lattice gauge theories on qudit
quantum computers},
\href{https://arxiv.org/abs/2507.12614}{arXiv:2507.12614} [quant-ph]
(2025).

\end{thebibliography}
\end{document}